\documentclass[jap,aps,twocolumn,showpacs,superscriptaddress]{revtex4}

\usepackage[]{graphicx}
\begin{document}



\title[Organic Single-Crystal Field-Effect Transistors]{Organic Single-Crystal Field-Effect Transistors}
\author{R. W. I. de Boer}
\affiliation{Kavli Institute of Nanoscience Delft, Delft
University of Technology, Lorentzweg 1, 2628 CJ Delft, the
Netherlands}

\author{M. E. Gershenson\footnote{Electronic mail:
gersh@physics.rutgers.edu}}
\affiliation{Department of Physics and
Astronomy, Rutgers University, Piscataway, New Jersey 08854, USA}

\author{A. F. Morpurgo\footnote{Electronic mail: A.Morpurgo@tnw.tudelft.nl}}
\affiliation{Kavli Institute of Nanoscience Delft, Delft
University of Technology, Lorentzweg 1, 2628 CJ Delft, the
Netherlands}

\author{V. Podzorov}
\affiliation{Department of Physics and Astronomy, Rutgers
University, Piscataway, New Jersey 08854, USA}

\begin{abstract}
We present an overview of recent studies of the charge transport
in the field effect transistors on the surface of single crystals
of organic low-molecular-weight materials. We first discuss in
detail the technological progress that has made these
investigations possible. Particular attention is devoted to the
growth and characterization of single crystals of organic
materials and to different techniques that have been developed for
device fabrication. We then concentrate on the measurements of the
electrical characteristics. In most cases, these characteristics
are highly reproducible and demonstrate the quality of the single
crystal transistors. Particularly noticeable are the small
sub-threshold slope, the non-monotonic temperature dependence of
the mobility, and its weak dependence on the gate voltage. In the
best rubrene transistors, room-temperature values of $\mu$ as high
as 15 cm$^2$/Vs have been observed. This represents an
order-of-magnitude increase with respect to the highest mobility
previously reported for organic thin film transistors. In
addition, the highest-quality single-crystal devices exhibit a
significant anisotropy of the conduction properties with respect
to the crystallographic direction. These observations indicate
that the field effect transistors fabricated on single crystals
are suitable for the study of the \textit{intrinsic} electronic
properties of organic molecular semiconductors. We conclude by
indicating some directions in which near-future work should focus
to progress further in this rapidly evolving area of research.
\end{abstract}

\maketitle

\renewcommand{\leftmark}
{R. W. I. de Boer \textit{et al.}: Organic Single-Crystal
Field-Effect Transistors}

\section{Introduction}

The electronic properties of Van-der-Waals-bonded organic
semiconductors are profoundly different from those of
covalently/ionically-bonded inorganic semiconductors
\cite{Silinsh94,Pope99}. In the highly-polarizable crystal
lattices of organic semiconductors, the electron-phonon coupling
is usually strong and the inter-molecular hopping amplitude small.
This results in the formation of self-trapped states with a size
comparable to the lattice constant, i.e., the small polarons. The
electronic, molecular and lattice polarization plays a key role in
determining transport in organic materials, as polaronic effects
"shape" both the \textit{dc} transport and optical properties of
these materials. Because of a very complicated character of the
many-particle interactions involved in polaron formation, this
problem has been treated mainly at the phenomenological level (see
Chapter 7 in Ref. \cite{Silinsh94}). Many basic aspects of this
problem have not been addressed yet, and a well-developed
microscopic description of the charge transport in organic
materials is still lacking.

Until recently, the experimental study of the low-frequency
\textit{intrinsic} electronic properties of organic semiconductors
have been performed only on \textit{bulk} ultra-pure crystals
\cite{Grosso01,Karl80}. In the time-of-flight (TOF) experiments by
the group of Norbert Karl at Stuttgart University
\cite{Karl85,Karl99}, it has been found that the mobility of
non-equilibrium carriers generated by light absorption in
ultra-high-purity oligomeric crystals can be as high as 400
cm$^2$/Vs at low temperatures (the latter $\mu$ value is
comparable to the mobility of electrons in Si MOSFETs at room
temperature). This behavior suggests that coherent, band-like
polaronic transport is possible in crystal of small organic
molecules.

To further investigate the electronic properties of organic
materials, it is important to go beyond the TOF measurements. One
of the alternative techniques to probe the charge transport on a
semiconductor surface is based on the electric field effect
\cite{Sze81}. Continuous tuning of the charge density induced by
the transverse electric field enables the systematic study of
charge transport, in particular the regime of large carrier
density that cannot be accessed in the TOF experiments. The field
effect forms the basis for operation of silicon field-effect
transistors (FETs), the workhorses of modern inorganic
electronics. The field-effect technique is also becoming
increasingly popular in the fundamental studies as a convenient
method to control the behavior of strongly correlated electron
systems such as high-temperature superconductors (see, e.g.,
\cite{Fiory90}) and colossal magnetoresistance manganites
\cite{Mathews97}. Other recent examples of applications of this
remarkably simple and very successful principle are the
electric-field tuning of the metal-insulator transition in
cuprates \cite{Newns98} and vanadium oxides \cite{Kim03}, and the
electrostatic control of ferromagnetism in Mn-doped GaAs
\cite{Ohno00}.

Organic semiconductors are, in principle, well suited for the
field-effect experiments. Owing to the weak van-der-Waals bonding,
the surface of organic semiconductors (e.g., polyacenes
\cite{Horowitz98,Katz00} and conjugated polymers \cite{Bao96}) is
characterized by an intrinsically low density of dangling bonds
that can act as the charge traps, and, hence, by a low threshold
for the field effect. This fact is at the origin of the rapid
progress of organic field-effect transistors based on thin film
technology, i.e., organic thin-film transistors (OTFTs)
\cite{Deleeuw00, Dimitrakopoulos02}.

Unfortunately, \textit{thin-film} transistors are \textit{not}
suitable for the study of \textit{intrinsic} electronic properties
of organic conductors, because their characteristics are often
strongly affected by imperfections of the film structure and by
insufficient purity of organic materials (see, e.g.,
\cite{Dimitrakopoulos02,Horowitz03,Campbell01}). As a consequence,
these devices commonly exhibit an exponential decrease of the
mobility of field-induced charge carriers with lowering
temperature \cite{Vissenberg98}. This behavior contrasts sharply a
rapid increase of $\mu$ with decreasing temperature, observed in
the TOF experiments with bulk ultra-pure organic crystals
\cite{Karl85,Karl99}. Because of a very strong dependence of the
OTFT parameters on fabrication conditions, some researchers came
to a pessimistic conclusion that even the best organic TFTs "may
not be appropriate vehicles for illuminating basic transport
mechanisms in organic materials" \cite{Nelson98}.

To explore the \textit{intrinsic} electronic properties of organic
materials and the physical limitations on the performance of
organic FETs, devices based on \textit{single-crystals} of organic
semiconductors are needed, similar to the single-crystal
structures of inorganic electronics. One of the major impediments
to realization of the single-crystal OFETs is the lack of
hetero-epitaxial growth technique for the Van-der-Waals-bonded
organic films. In this situation, the only viable option to study
the intrinsic charge transport on the surface of organic
semiconductors is to fabricate the field-effect structures on the
surface of free-standing organic molecular crystals (OMCs).
However, fabrication of single-crystal OFETs poses a technological
challenge. Because the surface of OMCs can be damaged much more
easily than that of their inorganic counterparts, organic
materials are by and large incompatible with conventional
microelectronic processing techniques such as sputtering,
photolithography, etc. This is why the systematic investigation of
single-crystal OFETs has been carried out only very recently
\cite{Podzorov03,Podzorov03a,DeBoer03,Takeya03,Butko03,Butko03a},
after the successful development of a number of novel fabrication
schemes (for earlier work see \cite{Horowitz96,Ichikawa02}).

Realization of the single-crystal OFETs opens a new avenue for the
study of charge transport in highly ordered molecular systems. The
use of single-crystal OFETs as an experimental tool enables the
investigation of aspects of charge transport in organic materials
that could not be addressed in the TOF experiments. One of the
important distinctions between these two types of experiments is
the magnitude of carrier densities. Very low densities of charge
carriers in the TOF experiments make interactions between them
insignificant. At the same time, in the field-effect experiments
with organic materials, where accumulation of $\sim 1$ carrier per
molecule seems to be feasible with the use of high-\textit{k}
dielectrics, these interactions could play a major role. Indeed,
it is well-known that at a sufficiently high density of
chemically-induced carriers, the potassium-doped fullerene
K$_{\mathrm{x}}$C$_{60}$ exhibits superconductivity (x $=3$) and a
Mott-Hubbard insulating state (x $=4$) \cite{Hebard91}. This
example illustrates a great potential of experiments with the
single-crystal OFETs.

The first working FET on the surface of a free-standing organic
molecular crystal has been fabricated a year ago
\cite{Podzorov03}. Though this field is in its infancy, the
progress has been remarkably rapid, with new record values of
carrier mobility for OFETs being achieved, new promising organic
materials being introduced, and new device processing techniques
being developed. In this review, we discuss the new techniques
responsible for the progress, the state-of-the-art characteristics
of single-crystal OFETs, and the experiments that show that
development of single-crystal OFETs enables investigation of
intrinsic electronic properties of organic materials. In the
future, the combined efforts of experimenters and theorists,
physicists and chemists will be required to reveal the full
potential of this research area. We hope that this paper will
provide a timely source of information for the researchers who are
interested in the progress of this new exciting field.

The structure of this review is as follows. In Section
\ref{fabrication}, we present an overview of the organic crystal
growth and techniques for the single-crystal OFET fabrication,
with a special attention paid to the crystal characterization.
Discussion of the single-crystal OFET characteristics in Section
\ref{characterization} focuses on revealing the intrinsic
transport properties of organic semiconductors. Finally, in
Section \ref{conclusion}, we summarize the main results, and
attempt to predict the directions of rapid growth in this
fascinating research field.

\section{Fabrication of single-crystal organic FETs} \label{fabrication}

The successful realization of FETs on the surface of organic
molecular crystals (OMC) is an important milestone in the research
of electronic transport in organic semiconductors. For the first
time, it opens the opportunity to study the intrinsic behavior of
charges at the organic surface, not limited by structural defects.
Fabrication of single organic crystal FETs comprises two main
steps: the growth of an organic crystal with atomically-flat
surface and preparation of the field-effect structure on this
surface. In this section we discuss both aspects, paying special
attention to the crystal characterization and to the analysis of
advantages and limitations of different fabrication methods.

\subsection{Single-crystal growth} \label{growth}

\begin{figure}[t]
\centering
\includegraphics[width=8.5cm]{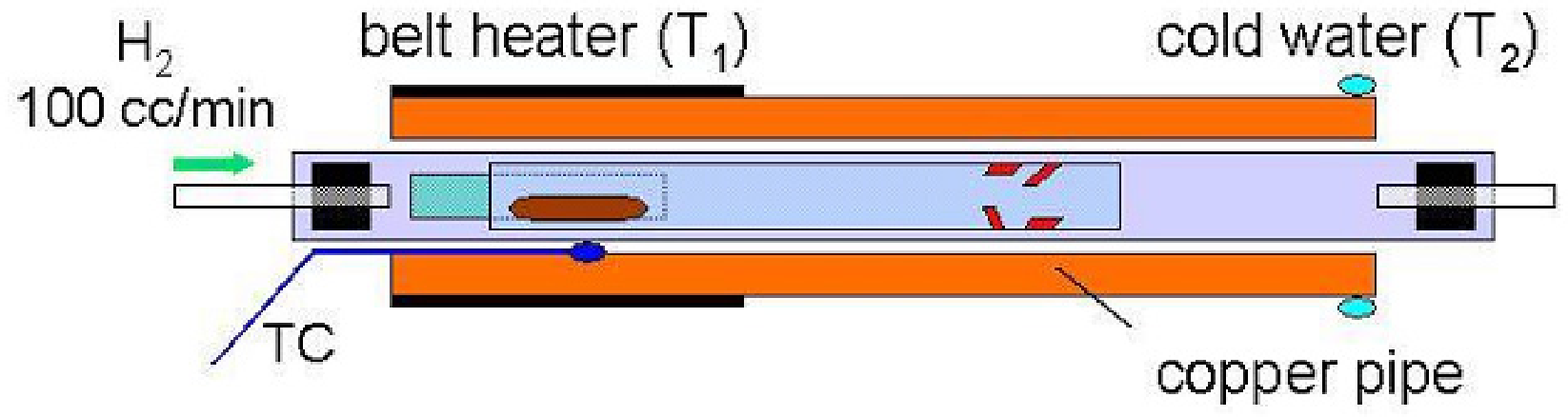}
\includegraphics[width=8.5cm]{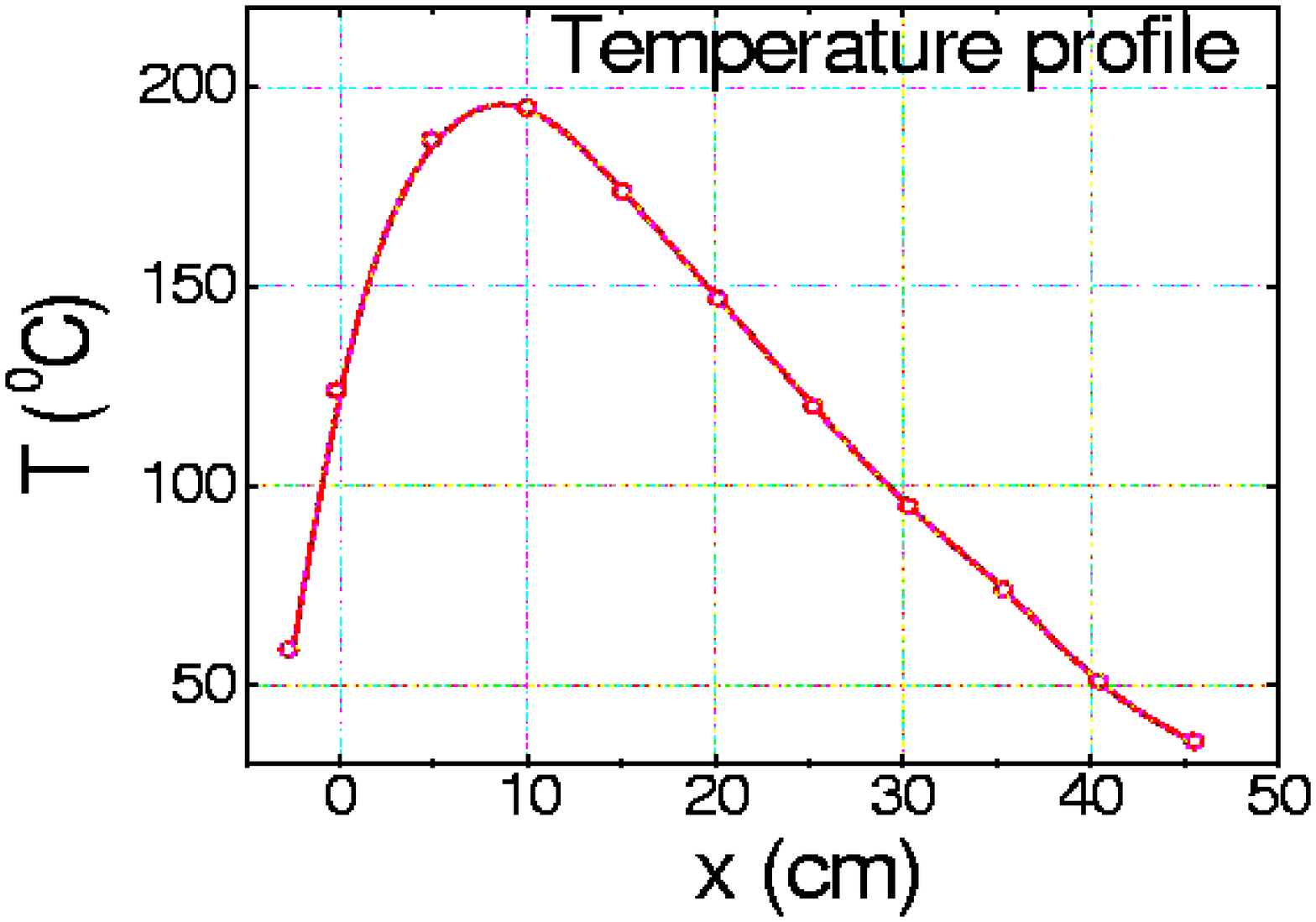}
\caption{Schematic overview of crystal growth system. Organic
material sublimes at temperature $T_1$, is transported through the
system by the carrier gas and recrystallizes in the cooler end of
the reactor. Heavy impurities (with a vapor pressure lower that
that of the pure organic compound) remain at the position of the
source material. Light impurities (with a vapor pressure higher
than that of the pure organic compound) condense at a lower
temperature, i.e. at a different position from where the crystals
grow. Therefore, the crystal growth process also results in the
purification of the material. \label{furnaceschematic}}
\end{figure}

Most of the single crystals used so far for the fabrication of
organic FETs have been grown from the vapor phase in a stream of
transport gas, in horizontal reactors (glass or, better, quartz
tube) \cite{Kloc97,Laudise98} (for a notable exception, in which
the crystals have been grown from solution, see Ref.
\cite{Mas04}). In the Physical Vapor Transport (PVT) method, the
starting material is placed in the hottest region of the reactor,
and the crystal growth occurs within a narrow temperature range
near its cold end (see Fig. \ref{furnaceschematic}). For better
separation of larger and, presumably, purer crystals from the rest
of re-deposited material along the tube, the temperature gradient
should be sufficiently small (typically, $2-5^{\mathrm{o}}$C/cm).

\begin{figure}[t]
\centering
\includegraphics[width=8.5cm]{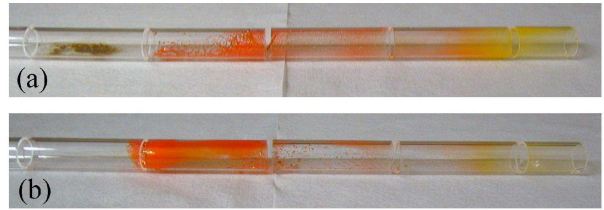}
\caption{(a) Result after first regrowth of as-purchased organic
material. Purified crystals are visible in the middle; the dark
residue present where the source material initially was and the
light (yellow) material visible on the right are due impurities.
(b) At the end of the second regrowth no dark residue is present
at the position of the source material, which demonstrate the
purifying effect of the growth process. \label{growthpictures}}
\end{figure}

Several ultra-high-purity gases have been used as a carrier agent:
in Ref. \cite{DeBoer03}, the highest mobility of tetracene-based
devices was realized with argon, whereas the best rubrene FETs
fabricated so far have been grown in pure H$_2$
\cite{Podzorov03,Podzorov03a}. In the latter case, hydrogen has
been chosen after comparison of the field-effect characteristics
of rubrene crystals grown in Ar, N$_2$, and H$_2$ atmospheres. It
is unclear at present how exactly the transport gas affects the
crystal quality; uncontrollable variations of the crystal quality
might be caused by the residual water vapor and oxygen in the
reactors. Photo-induced reactions with O$_2$ are known for most
organic molecules \cite{Dabestani96} and the products of these
reactions can act as traps for charge carriers. To minimize
possible photo-activated oxidation of organic material, the
reactors should be pumped down to a reduced pressure $P \simeq
10^{-2}$ mbar prior to the crystal growth, and the growth should
be performed in the dark.

Several factors affect the growth process and the quality of the
crystals. Important parameters are, for instance, the temperature
in the sublimation zone, $T_{\mathrm{sblm}}$ and the gas flow
rate. Many other factors can also play a role: e.g., acoustical
vibrations of the reactor in the process of growth might affect
the size, shape, and quality of the crystals. For each material
and each reactor, the optimal parameters have to be determined
empirically. At least in one case (the rubrene-based OFETs
\cite{Podzorov03a}), it has been verified that the slower the
growth process, the higher the field-effect mobility. For this
reason, the temperature of sublimated organic material was chosen
close to the sublimation threshold. The crystal growth in this
regime proceeds by the flow of steps at a very low rate ($\leq 5
\times 10^{-7}$ cm/s in the direction perpendicular to the
\textit{a-b} plane), and results in a flat surface with a low
density of growth steps \cite{Laudise70}. As an example,
sublimation of 300 mg of starting material in Ref.
\cite{Podzorov03a} took up to 48 hours at $T_{\mathrm{sblm}} =
300^{\mathrm{o}}$C.

Another important parameter is the purity of the starting
material. As the crystal growth process also results in the
chemical purification of the material, several re-growth cycles
may be required for improving the field-effect mobility, with the
grown crystals used as the starting material for the subsequent
re-growth. The number of required re-growth cycles depends
strongly on the purity of starting material. Figure
\ref{growthpictures} illustrates the need for several re-growth
cycles in the process of the growth of tetracene crystals. Despite
the nominal $98 \%$ purity of the starting tetracene
(Sigma-Aldrich), a large amount of residue left in the sublimation
zone after the first growth cycle is clearly visible (Fig.
\ref{growthpictures}a); this residue is not present at the end of
the second growth cycle (fig. \ref{growthpictures}b). A word of
caution is appropriate: in the authors' experience, different
batches of as-purchased material, though being of the same nominal
purity, might leave different amount of residue. Clearly, the
better purity of the starting material, the fewer re-growth cycles
are required for a high FET mobility: in Ref. \cite{Podzorov03a}
the rubrene OFETs with $\mu > 5$ cm$^2$/Vs have been fabricated
from the "sublimed grade" material (Sigma-Aldrich) after only $1 -
2$ growth cycles.

\begin{figure}[t]
\centering
\includegraphics[width=8.5cm]{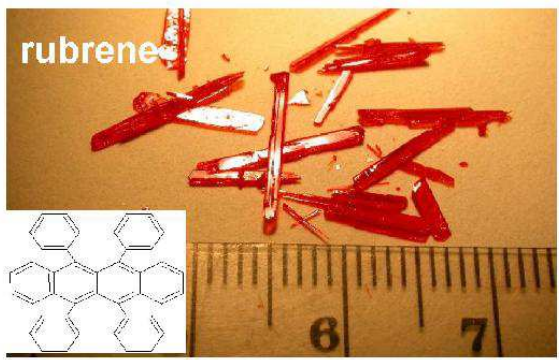}
\caption{Result of rubrene crystal growth. Most of the organic
crystals grown by the physical vapor transport are shaped as
elongated "needles" or thin platelets. \label{rubrenecrystals}}
\end{figure}

\begin{figure}[t]
\centering
\includegraphics[width=8.5cm]{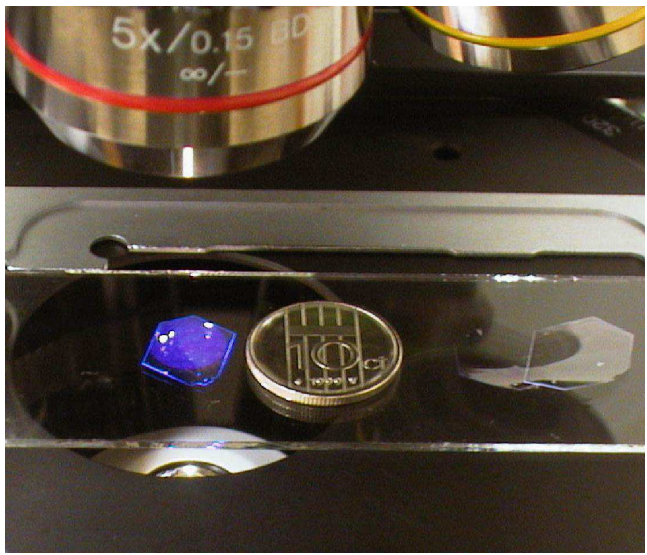}
\caption{Two platelet-shaped anthracene single-crystals grown by
physical vapor transport. The crystals are transparent and
colorless. The left crystal is illuminated by UV light, and
fluoresces in the blue. \label{anthracene}}
\end{figure}

\begin{figure}[t]
\centering
\includegraphics[width=8.5cm]{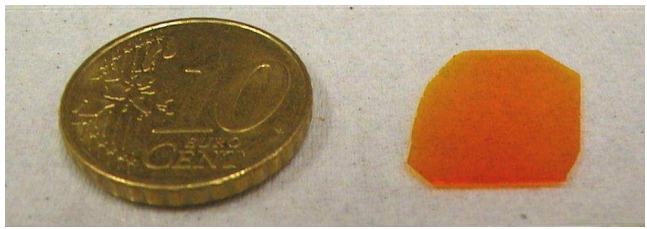}
\caption{A cm$^2$-sized platelet-shaped tetracene single-crystal
grown by physical vapor transport. \label{tetracenebig}}
\end{figure}

It is likely that the purity of crystals for the OFET fabrication
can be substantially improved if a zone refining process
\cite{Grosso01,Karl80} is used for pre-purification of the
starting material. Indeed, in the time-of-flight studies of
organic crystals, the best results and the highest mobilities have
been obtained after multiple zone-refinement purification cycles
\cite{Karl80,Karl85}. This process enabled reduction of the
impurity concentration in the bulk down to the part-per-billion
level. It is unlikely that a comparable purity can be achieved
simply by multiple vapor transport re-growth processes. It has to
be noted that zone-refinement cannot be applied to all organic
materials, since this technique requires the existence of a
coherent liquid phase (i.e. the melting temperature of the
substance has to be lower than the temperature of decomposition of
its molecules) \cite{Grosso01}. However, for several materials
that have already been successfully used for fabrication of single
crystal FETs (e.g., rubrene, perylene, anthracene), a coherent
liquid phase does exist and zone refinement is possible. The
zone-refining purification might be especially useful for
realization of intrinsic polaronic transport at low temperatures,
where trapping of polarons by defects becomes a serious problem.

Most of the organic crystals grown by the physical vapor transport
are shaped as elongated "needles" or thin platelets (see Fig.
\ref{rubrenecrystals}). The crystal shape is controlled by the
anisotropy of inter-molecular interactions: for many materials, a
larger crystal dimension corresponds to the direction of the
strongest interactions and, presumably, the strongest overlap
between $\pi$-orbitals of adjacent molecules. For this reason, the
direction of the fastest growth of needle-like rubrene crystals
coincides with the direction of the highest mobility of
field-induced carriers (see Sec. \ref{characterization}). For
platelet-like crystals, the larger facets are parallel to the
\textit{a-b} plane. Typical in-plane dimensions range from a few
square millimeters for rubrene to several square centimeters in
the case of anthracene. The crystal thickness also varies over a
wide range and, in most cases, can be controlled by stopping the
growth process at an early stage. For example, the thickness of
the tetracene crystals grown for 24 hours ranges between $\sim 10
\ \mu$m and $\sim 200 \ \mu$m \cite{DeBoer04}, but it is possible
to harvest several crystals of sub-micron thickness by stopping
the growth process after $\sim 30$ minutes.

Because of a weak van der Waals bonding between the molecules,
polymorphism is a common phenomenon in organic materials: the
molecular packing and the shape of organic crystals can be easily
affected by the growth conditions. For example, the thiophenes
exhibit two different structures depending on the growth
temperature \cite{Horowitz95}. In many cases, organic molecular
crystals exhibit one or more structural phase transitions upon
lowering the temperature. For the study of single-crystal FETs at
low temperature, the occurrence of a structural phase transition
can be detrimental. In tetracene crystals, for instance, a
structural phase transition occurs below 200 K (see, e.g.,
\cite{Sondermann85}). Co-existence of two crystallographic phases
at lower temperatures causes the formation of grain boundaries and
stress, which are responsible for the trapping of charge carriers
(see Fig. \ref{SCLCmobvstemp}). In tetracene, in addition,
occurrence of the structural phase transition often results in
cracking of the crystals with cooling and a consequent device
failure.

\subsection{Crystal characterization} \label{crystalchar}

To understand better the effect of different factors on the
crystal growth, a thorough characterization of the crystal
properties is needed. Note, however, that many experiments provide
information on the crystal properties that is only indirectly
related to the performance of the single-crystal OFETs. For
example, the x-ray analysis of organic crystals, though necessary
for identification of the crystal structure and orientation of the
crystallographic axes, is insufficiently sensitive for detection
of a minute concentration of defects that might severely limit the
field-effect mobility at low temperatures. Similarly, the TOF
experiments, although useful in assessing the quality of the {\it
bulk} of organic crystals, are not sensitive to the surface
defects that limit the OFET performance. Below we briefly review
several techniques that have been used for organic crystal
characterization.

\subsubsection{Polarized-light microscopy}

Inspection of crystals under an optical microscope in the
polarized light provides a fast and useful analysis of the
crystalline domain structure. Visualization of domains is possible
because crystals of most organic conjugated materials are
birefringent \cite{Vrijmoeth98}. Optical inspection also enables
detection of the stress in crystals, which results in appearance
of the interference fringes with orientation not related to any
specific crystallographic direction. This technique simplifies the
process of selection of single crystals for transport
measurements.

\subsubsection{The time-of-flight experiments}

In the time-of-flight (TOF) experiments, a platelet-like crystal
is flanked between two metal electrodes, one of which is
semi-transparent \cite{Karl85}. A thin sheet of photo-excited
charge carriers is generated near the semi-transparent electrode
by a short laser pulse with the photon energy greater than the
band gap. In the presence of a constant voltage bias between the
two electrodes, the charge sheet propagates in the direction
determined by the \textit{dc} electric field and generates a
displacement current, whose magnitude diminishes rapidly as soon
as the sheet reaches the opposite electrode \cite{Grosso01}. From
the duration of the displacement current pulse and the known
crystal thickness, the drift velocity and carrier mobility can be
calculated. This method also provides indirect information on the
concentration of (shallow) traps in the bulk: the decrease of
mobility at low temperatures is caused by multiple trapping and
release processes (for more details, see Ref. \cite{Hoesterey63}).
An important aspect of TOF measurements is that their results are
not sensitive to contact effects, since the charge carriers are
photo-generated (i.e., not injected from a metal electrode) and
their motion is detected capacitively. This simplifies the contact
preparation and improves the reproducibility of results for
identically grown crystals.

\begin{figure}[b]
\centering
\includegraphics[width=8.5cm]{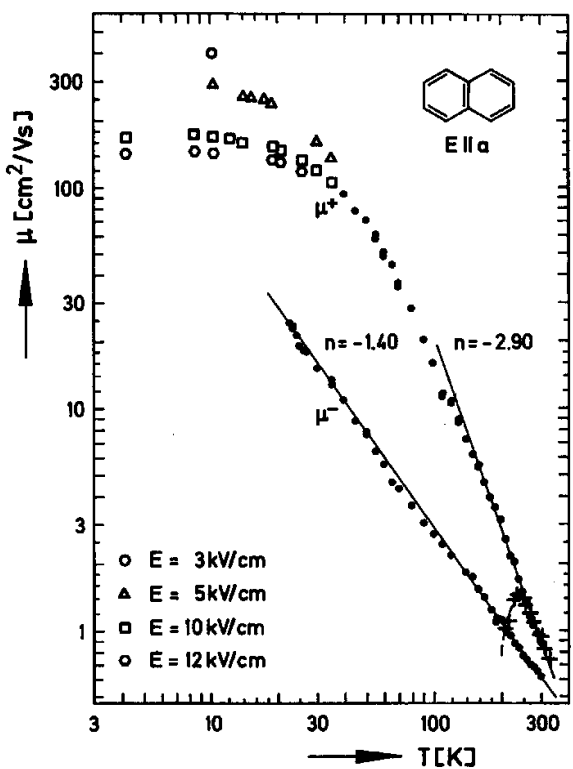}
\caption{Electron and hole mobility $\mu$ versus temperature $T$
in ultra-pure single crystals of naphthalene, as measured in
Time-Of-Flight (TOF) experiments. The solid lines indicate a
$T^{-n}$ power-law temperature dependence with exponents $n$ as
indicated in the figure. For holes, mobility values as high as 400
cm$^2$/Vs are observed at low temperature. (\cite{Karl85}, printed
with permission of N. Karl, Crystal Laboratory, Univ. Stuttgart.)
\label{Karlpicture}}
\end{figure}

For measuring the intrinsic mobility of charge carriers in the
bulk, the lifetime of the carriers against charge trapping has to
be greater than the time of flight between the electrodes. This
requirement imposes severe limitation on the concentration of
charge traps. As a result, only very pure and defect-free crystals
can be characterized by the TOF method. For example, according to
preliminary measurements by the Stuttgart group
\cite{Stuttgart04}, the rubrene crystals used for fabrication of
the high-mobility OFETs \cite{Podzorov03a} are unsuitable for the
TOF measurements. The crystals for the TOF measurements should
also have sufficiently parallel opposite facets and be
sufficiently thick for the displacement current pulse to be longer
than the apparatus time resolution. Because of these limitations,
the TOF measurements can be performed only on a small fraction of
the crystals grown by the phase vapor deposition technique.

Despite the difficulties of application of the TOF method to the
organic crystal grown by vapor transport, successful TOF
measurements have been performed on vapor-grown tetracene crystals
similar to those used in FETs experiments \cite{DeBoer04}. The
room-temperature mobility $\mu = 0.5 - 0.8$ cm$^2$/Vs, measured in
the TOF experiment for three different crystals, is comparable to
the highest mobility of the field-induced carriers in the OFET
experiments; these quantities also exhibit similar temperature
dependencies. Interestingly, the two types of measurements
provided similar $\mu$ values for charge transport along different
directions in anisotropic organic crystals: the FET measurements
probe surface transport in the \textit{a-b} plane, whereas TOF
experiments on the platelet tetracene crystals probed motion along
the \textit{c}-axis. This observation seems to be in disagreement
with what one would expect from the crystallographic structure of
the material, namely a pronounced anisotropy of mobility along
different crystallographic directions
\cite{Karl99,Karl85a,Karl01}. To better understand the origin and
the implications of this result, further characterization of a
larger variety of organic crystals by the TOF method and
comparison with field-effect measurements are needed.

\subsubsection{Space charge limited current spectroscopy}

A rather common and, in principle, simple way to characterize the
electrical properties of OMCs is the study of the $I-V$
characteristics measured in the space charge limited current
(SCLC) regime \cite{Lampert70}. The value of the carrier mobility,
as well as the density of deep traps can be inferred from these
measurements. Similar to the TOF experiments, SCLC measurements
require relatively large electric fields. Usually, these
measurements are performed on thin platelet-like crystals; the
metallic contacts are located on the opposite facets of a crystal
of thickness $L$, and the current $I$ in the direction
perpendicular to the surface (along the \textit{c}-axis) is
measured as a function of the voltage $V$ between the contacts.
Measurements of $I-V$ characteristics with contacts on the same
surface of the crystal can also be performed, but in this geometry
the extraction of the carrier mobility and the density of traps
from the experimental data is more involved. Instrumentation-wise,
the \textit{dc} charge injection experiments are less challenging
than the TOF measurements. However, a high quality of contacts is
required for the former measurements, otherwise the results for
nominally identical samples are not reproducible \cite{DeBoer04}.
Because of a high sensitivity of the data to the contact quality,
the SCLC technique typically requires acquisition of a large
volume of data for many samples and, therefore, is not very
efficient.

\begin{figure}[t]
\centering
\includegraphics[width=8.5cm]{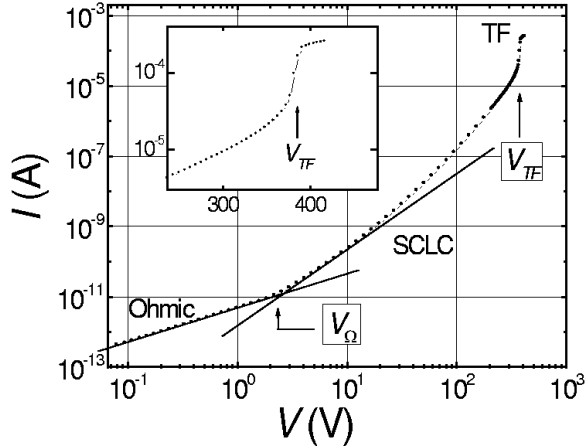}
\caption{$I-V$ characteristic of a $\sim 10 \ \mu$m-thick rubrene
crystal, measured along the \textit{c}-axis. The inset is a
blow-up of the crossover to the trap free regime (also in a
double-log scale). From the crossover to the trap-free regime, the
density of deep traps $N_{t}^{d} \simeq 10^{15}$ cm$^{-3}$ can be
estimated \cite{Podzorov03a}. \label{SCLCrubrene}}
\end{figure}

\begin{figure}[t]
\centering
\includegraphics[width=8.5cm]{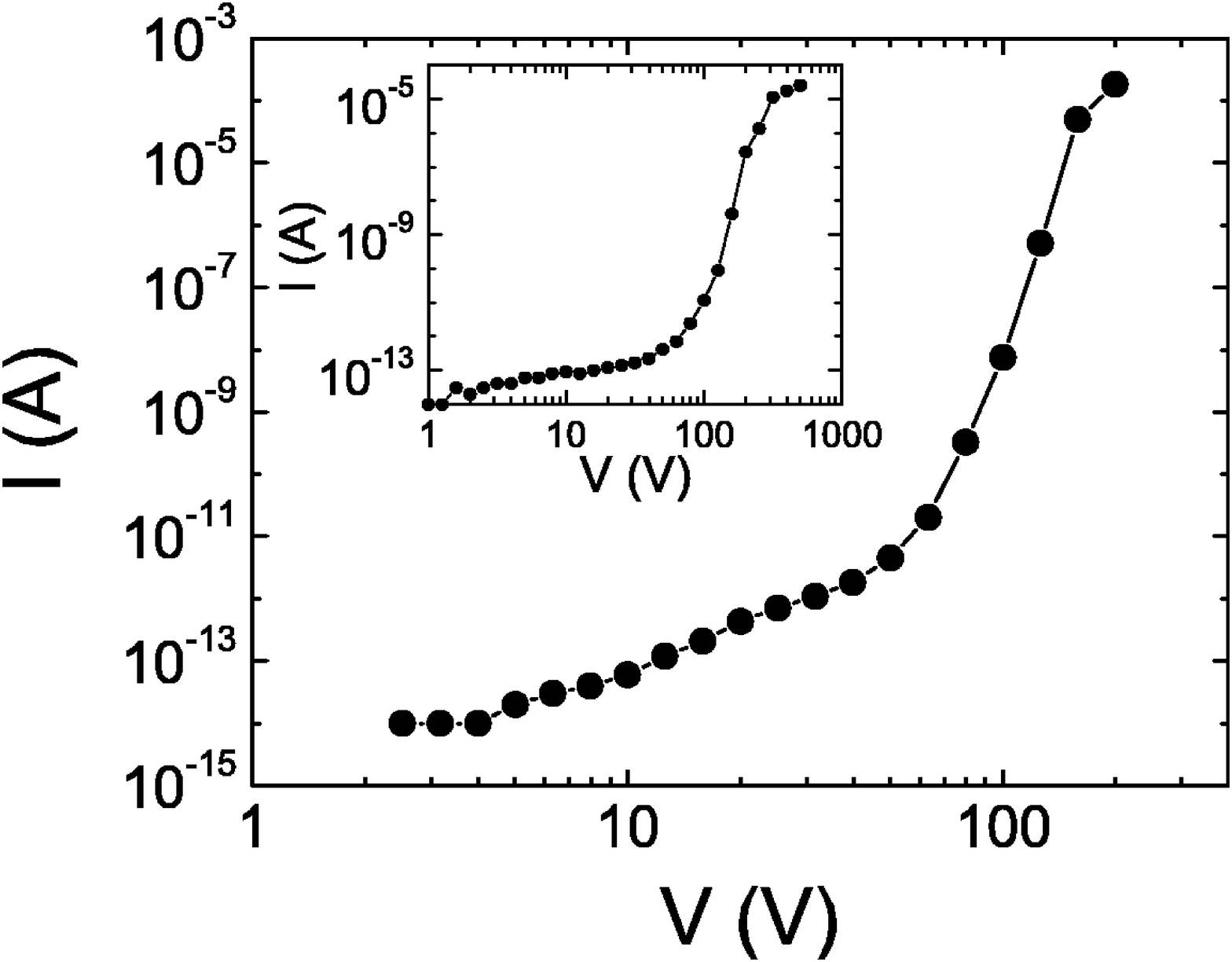}
\caption{Typical result of a DC $I-V$ measurement perpendicular to
the \textit{a-b} plane of a tetracene single-crystal, with a
thickness $L = 30 \ \mu$m and a mobility $\mu_{\mathrm{min}} =
0.59$ cm$^2$/Vs. The inset shows a similar measurement on a
different crystal ($L = 25 \ \mu$m, $\mu_{\mathrm{min}} = 0.014$
cm$^2$/Vs), in which a crossing over into an approximately
quadratic dependence on voltage is visible at high voltage. In
both cases, a very steep current increase occur around of just
above 100 V that we attribute to filling of deep traps. We
observed a steep increase in current in most samples studied.
\label{SCLCtypical}}
\end{figure}

Recently, the trap-free space charge limited current (SCLC) regime
has been observed for samples with both thermally-evaporated
thin-film contacts \cite{Podzorov03a} and silver epoxy contacts
\cite{DeBoer04}. The $I-V$ characteristics for a thin rubrene
crystal with thin-film silver contacts demonstrate the crossovers
from the Ohmic regime to the space-charge-limited-current (SCLC)
regime, and, with a further voltage increase, to the trap-free
(TF) regime (see Fig. \ref{SCLCrubrene}) \cite{Podzorov03a}.
Observation of a linear Ohmic regime indicates that the non-linear
contribution of Schottky barriers formed at the metal/rubrene
interfaces is negligible in these measurements: the voltage drop
across the Schottky barriers is smaller than the voltage drop
across the highly resistive bulk of the crystal. The crossover to
the SCLC regime at a low bias voltage $V_{\Omega} \simeq 2.5$ V
suggests that the charge carrier injection from the contacts is
very efficient. From the threshold voltage of the TF regime,
$V_{TF}$, the density of deep traps, $N_{t}^{d}$, can been
estimated \cite{Lampert70,Kao81}:
\begin{equation}
N_{t}^{d} = \frac{\epsilon \epsilon_{0} V_{TF}}{e L^2}
\end{equation}
Here $\epsilon_{0}$ is the permittivity of free space, $\epsilon$
is the relative dielectric constant of the material. Note that an
assumption-free estimate of the trap density can be made only if a
well-defined crossover between SCLC and TF regimes is observed.
For this reason, the method can be applied only to sufficiently
pure crystals. For the tetracene crystals studied in Ref.
\cite{DeBoer04}, $N_t \simeq 5 \times 10^{13}$ cm$^{-3}$ is
significantly smaller than $N_t \simeq 10^{15}$ cm$^{-3}$ for
rubrene crystals studied in Ref. \cite{Podzorov03a}. The steep
increase of current that signifies the transition to the trap-free
regime, is also much more pronounced for tetracene crystals (see
Fig. \ref{SCLCrubrene} and \ref{SCLCtypical}).

The estimate of $N_{t}^{d}$ is based on the assumption (not
usually mentioned in the literature), that the deep traps are
uniformly distributed throughout the entire crystal bulk. However,
it is likely that the trap density is greater near the
metal/organic interface because of the surface damage during the
contact preparation. A small amount of traps located close to the
surface can have a large effect on the current flow: the charges
trapped near the surface strongly affect the electric field in the
bulk of the crystal, which determines the current flow in the TF
regime. For this reason, the value of $N_{t}^{d}$ may be
considered as an upper limit of the actual density of traps in the
bulk (see Ref. \cite{DeBoer04} for a more detailed discussion).

\begin{figure}[b]
\centering
\includegraphics[width=8.5cm]{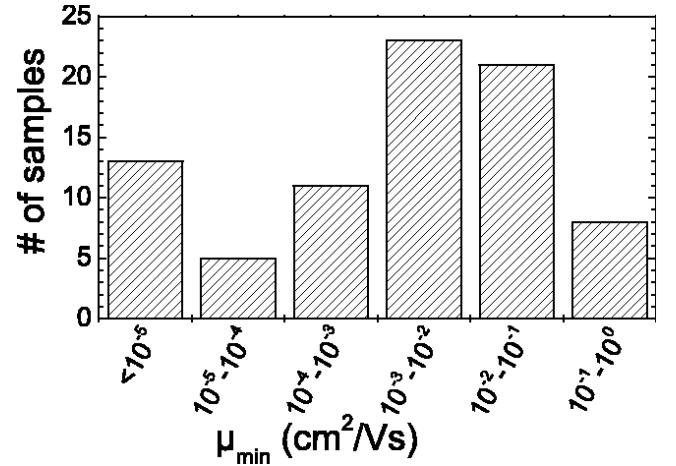}
\caption{Histogram of values for $\mu_{\mathrm{min}}$ calculated
from \textit{dc} $I-V$ measurements performed on approximately 100
tetracene single-crystals. The large scattering in the observed
values is due to the spread in contact quality.
\label{SCLCmobilitiesoverview}}
\end{figure}

In the TF regime, the mobility can be estimated from the
Mott-Gurney law for the trap-free regime \cite{Lampert70}:
\begin{equation}
J_{TF} = \frac{9 \epsilon \epsilon_{0} \mu V^{2}}{8 L^{3}}
\end{equation}
where $J_{TF}$ is the current density. Even when the TF limit is
not experimentally accessible, the same formula can be used to
extract a lower limit, $\mu_{\mathrm{min}}$ for the intrinsic
mobility, at least in materials in which one type of carriers
(electrons or holes) is responsible for charge transport (see Ref.
\cite{DeBoer04} for details). In tetracene crystals, the values of
$\mu_{\mathrm{min}}$ along the \textit{c}-axis obtained from SCLC
measurements performed on a large number of samples are shown in
Fig. \ref{SCLCmobilitiesoverview}. The large spread in values for
$\mu_{\mathrm{min}}$ obtained from identically grown crystals is
due to scattering in the contact parameters.

\begin{figure}[b]
\centering
\includegraphics[width=8.5cm]{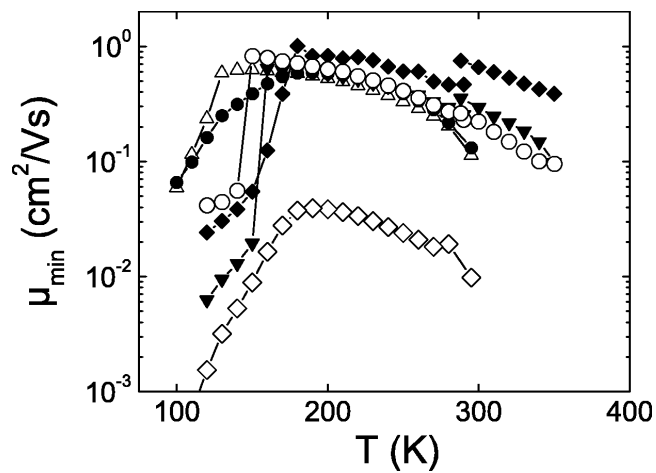}
\caption{Temperature dependence of the lower limit to the
mobility, $\mu_{\mathrm{min}}$, measured for tetracene
single-crystals with a high $\mu_{\mathrm{min}}$ value. Note the
abrupt drop in mobility occurring at different temperatures below
$\simeq 180$ K, originating from a known structural phase
transition \cite{Sondermann85}. \label{SCLCmobvstemp}}
\end{figure}

For the samples with $\mu_{\mathrm{min}} > 0.1$ cm$^2$/Vs, the
mobility increases with cooling over the interval $T \simeq 180 -
300$ K (Fig. \ref{SCLCmobvstemp}). Observation of the mobility
increase with cooling in high-quality crystals is usually
considered as a signature of the intrinsic (disorder-free)
transport (for comparison, an increase of mobility with lowering
temperature has never been observed in SCLC measurements performed
on disordered thin organic films). With further cooling, however,
the mobility decreases: Fig. \ref{SCLCmobvstemp} shows that for
most of the tetracene crystals, a sharp drop of $\mu$ below 180 K
is observed. This suggests that this drop might be related to a
structural phase transition, which is known to occur in tetracene
in this temperature range \cite{Sondermann85}. Both the observed
effects of the structural phase transition on the carrier
mobility, as well as the increase in $\mu_{min}$ with lowering
temperature, indicate that these SCLC measurements reflect the
intrinsic electronic properties of the material.

\subsection{Fabrication of the field-effect structures} \label{FETstructures}

Fabrication of the field-effect structure on the surface of
organic crystals poses a challenge, because many conventional
fabrication processes irreversibly damage the surface of
van-der-Waals-bonded organic crystals by disrupting molecular
order, generating interfacial trapping sites, and creating
barriers to charge injection. For example, sputtering of an
insulator onto the crystal creates such a high density of defects
on the organic surface that the field-effect is completely
suppressed. Up to date, two techniques for single-crystal OFET
fabrication have been successfully used: (a) electrostatic
"bonding" of an organic crystal to a prefabricated
source/drain/gate structure, and (b) direct deposition of the
contacts and gate insulator onto the crystal surface. In this
section, we address the technical aspects of these fabrication
processes; effect of different fabrication methods on the
electrical characteristics of the resulting single-crystal FETs
will be discussed in Sec. \ref{characterization}.

\subsubsection{Electrostatic bonding technique} \label{adhesion}

In this approach, the transistor circuitry (both gate and
source/drain electrodes) is fabricated on a separate substrate,
which, at the final fabrication stage, is electrostatically bonded
to the surface of organic crystal. The technological processes
vary depending on the type of a substrate for the transistor
circuitry. Two kinds of substrates have been used: conventional
silicon wafers \cite{DeBoer03,Takeya03,Ichikawa02}, and flexible
elastomer substrates (the so-called rubber stamps)
\cite{Rogers03}.

\paragraph{Source/drain/gate structures on Si substrates}

\begin{figure}[t]
\centering
\includegraphics[width=8.5cm]{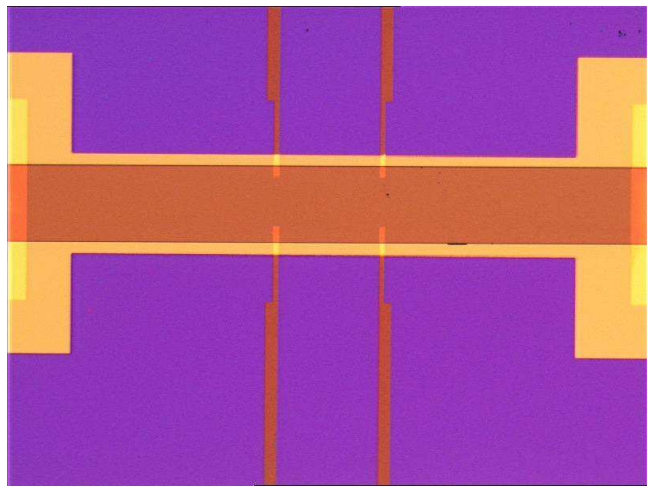}
\caption{Optical microscope picture of a single-crystal rubrene
FET, fabricated by electrostatic bonding. The crystal, which has a
rectangular shape, overlaps with the source and drain contacts (at
the left and right edge of the picture) and with four small
contacts in the center, used to perform 4-probe electrical
measurements. The purple area consists of a Ta$_2$O$_5$ layer
sputtered on top of the substrate prior to the crystal bonding,
which, for wider crystals, serves to confine the electrically
active region of the FET. \label{rubreneFETpicture}}
\end{figure}

In this method, the source/drain/gate structure is fabricated on
the surface of a heavily doped (n-type or p-type) Si wafer,
covered with a layer of thermally grown SiO$_2$ (typically, $0.2 \
\mu$m thick). The conducting Si wafer serves as the gate
electrode, and the SiO$_2$ layer plays the role of the gate
insulator. The source and drain gold contacts are deposited on top
of the SiO$_2$ layer, and, as a final step, a sufficiently thin
OMC crystal is electrostatically bonded to the source/drain/gate
structure. It has been found in Ref. \cite{DeBoer03} that the
reactive ion etching (RIE) of the contact/SiO$_2$ surface in the
oxygen plasma prior to the OMC bonding improves significantly the
characteristics of tetracene OFETs: the RIE cleaning reduces the
spread of mobilities, the field-effect threshold voltage, and the
hysteresis of transfer characteristics. The RIE cleaning also
significantly improves adhesion of freshly grown tetracene
crystals to SiO$_2$ surface. The technique works best for very
thin crystals ($\leq 1 \ \mu$m thick) that adhere spontaneously to
the substrate, but it can also be applied (with a lower success
yield) to much thicker crystals by gently pressing on the crystal
to assist the adhesion process \cite{Jurchescu03}. Fig.
\ref{rubreneFETpicture} shows a picture of a rubrene FET
fabricated with the technique of electrostatic adhesion to
SiO$_2$.

\paragraph{Source/drain/gate structures on flexible substrates}

\begin{figure}[t]
\centering
\includegraphics[width=8.5cm]{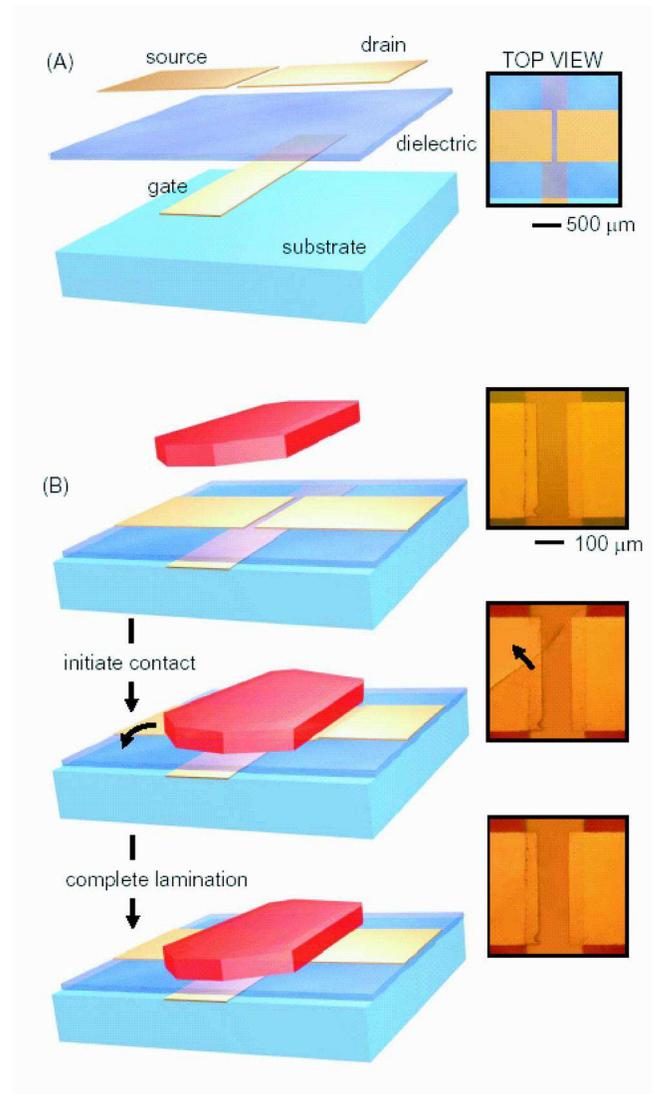}
\caption{The PDMS-based stamp: source/drain/gate structures on a
flexible substrate. (a) Schematic picture of the layout of the
stamp. It consists of a few-mm-thick PDMS pad, a patterned gate
electrode, a (2-4)$\mu$m-thick PDMS spin-coated film as a
gate-dielectric, and gold source and drain electrodes. (b) The
principle of lamination. The crystal is positioned on the stamp
surface, and a gentle pressure is applied to one edge of the
crystal. Van-der-Waals forces then spontaneously cause a "wetting"
front to proceed across the crystal surface. The insets at the
right show microscope pictures of the transparent crystal on top
of the stamp at different stages of the lamination process.
\label{stampschematic}}
\end{figure}

In this approach, the FET circuitry is fabricated on top of a
flexible elastomer substrate (polydimethylsiloxane, or PDMS), by
sequential shadow-mask deposition of the gate and source/drain
electrodes \cite{Rogers03}. The fabrication process, illustrated
in Fig. \ref{stampschematic}, begins with deposition of the gate
electrode on top of a few-mm-thick PDMS substrate (1.5 nm of Ti as
an adhesion promoter to the substrate, 20 nm of Au, and 3 nm of Ti
as an adhesion promoter to the subsequent layers). A
(2-4)$\mu$m-thick PDMS film, deposited by spin-coating on top of
the structure, serves as a gate dielectric. Evaporation or
transfer printing of source and drain electrodes (1.5 nm of Ti and
20 nm of Au) on top of the dielectric completes the stamps.
Careful control of the fabrication processes results in electrode
and dielectric surfaces with low surface roughness (the
root-mean-squared value of $\sim 0.6$ nm, as measured by atomic
force microscopy). The final assembly of the devices, similar to
the Si-based technique, consists of positioning of the OMC crystal
on the stamp surface, and applying a gentle pressure to one edge
of the crystal. Van der Waals forces then spontaneously cause a
"wetting" front to proceed across the crystal surface at a rate of
a few tenths of a millimeter per second. This lamination process
yields uniform contact, devoid of air gaps, bubbles or
interference fringes.

The main advantage of both Si- and PDMS-based stamps technique is
obvious: it eliminates the need for deposition of metals and
dielectrics directly onto a very vulnerable organic surface. Since
these techniques exploit the technologies well-developed in
electronic industry, the dimension of the circuitry can be easily
reduced, if desirable, well in the sub-micron range: specifically,
a very small source/drain contact separation can be achieved by
using electron-beam lithography. The PDMS-based masks work well
not only for thin, but also for thick crystals: the flexible PDMS
surface and the ductile Au contacts adjust easily to the crystal
shape (i.e. no flexibility of the crystal is required).
Remarkably, despite the fact that rubbers become rigid upon
cooling, this technique has been shown to work well at low
temperatures \cite{Sundar04}. Another important advantage of the
PDMS stamp technique is that it is non-destructive and
\textit{reversible}: it has been shown that the contact between
the stamp and the rubrene crystal can be re-established many times
without noticeable degradation of the OMC surface \cite{Rogers03}.
For this reason, PDMS-supported circuitry has been used for the
first observation of the anisotropy of the field-effect mobility
within the \textit{a-b} plane of the single crystals of rubrene,
as we discuss in Sec. \ref{anisotropy}. Interestingly, simple
adhesion of organic crystals to metallic surface results in
contacts with good electrical properties. This has been
demonstrated for both Si- and PDMS- based stamps by comparing the
results of two- and four-probe measurements. The contact
resistance is similar to that observed in the devices for which
the thin metal film contacts are directly deposited onto the
organic surface (see below).

Although its simplicity makes electrostatic bonding particularly
appealing for the fabrication of organic single crystal FETs, this
technique also suffers from a number of limitations. For instance,
the choice of metals for the source and drain contacts in the
electrostatic "bonding" technique is limited by noble metals,
since other materials are easily oxidized in air. The channel
width is not well defined (unless it is limited by the crystal
dimensions or by patterning the gate electrode), because the
conduction channel is formed over the whole area of overlap
between the OMC crystal and the Si wafer. In the current
PDMS-stamp OFETs, the gate insulator is relatively thick and its
dielectric constant is low - for this reason, the maximum density
of induced charges is relatively small (typically, below $1 \times
10^{11}$ carriers/cm$^2$). Finally, another potential problem of
the Si- and PDMS stamps might be the mismatch between the
coefficients of thermal expansion for the stamp and the organic
crystal. Upon changing the temperature, this mismatch might cause
a mechanical stress and formation of defects in the crystal. Since
the surface defects can trap the field-induced charge, this might
result in deterioration of the low-temperature OFET
characteristics. This is an important issue that requires further
studies.

\paragraph{OFETs with high-\textit{k} dielectrics}

The electrostatic bonding technique is compatible with the use of
high-\textit{k} dielectrics as gate insulators, which allow the
accumulation of a large carrier density in OFETs. Particularly
interesting is the possibility of reaching a surface charge
density of the order of $1 \times 10^{14}$ carriers/cm$^2$, which
corresponds to approximately one charge carrier per molecule (this
estimate assumes that all the charges are accumulated in a single
molecular layer, as it is expected from calculations of the
screening length). Indeed, the maximum surface charge density is
$Q = \epsilon \epsilon_0 E_{bd}$, where $\epsilon$ is the relative
dielectric constant of the gate dielectric and $E_{bd}$ is its
breakdown field. For typical high-\textit{k} dielectrics, such as
Ta$_2$O$_5$ or ZrO$_2$, $\epsilon = 25$ and $E_{bd} > 6$ MV/cm,
and the resulting charge density at the breakdown is $10^{14}$
carriers/cm$^2$. Many novel high-\textit{k} materials hold the
promise of even higher charge densities \cite{Dover98,Ren01}.

The process of fabrication of OFETs with high-\textit{k}
dielectrics is similar to the aforementioned Si-based technique,
with SiO$_2$ replaced by a high-\textit{k} dielectric that is
sputtered onto a heavily doped silicon substrate. In the
experiments at TU Delft, Ta$_2$O$_5$ and ZrO$_2$ gate dielectrics
have been sputtered onto the Si wafers at room temperature. Though
the measured values of the dielectric constant and breakdown field
for these layers are close to the best results reported in
literature \cite{Fleming00}, the leakage currents were relatively
large ($\sim 10^{-6}$ A/cm$^2$ for a Ta$_2$O$_5$-thickness of 350
nm). This is typical for deposition of dielectric films onto
non-heated substrates, which results in a relatively high density
of vacancies in the films. Substantially lower values of the
leakage currents might be achieved in the future by sputtering on
heated substrates, as already demonstrated in literature
\cite{Fleming00}. The characteristics of tetracene single-crystal
FETs with high-\textit{k} dielectrics will be discussed in Sec.
\ref{characterization}.

\subsubsection{"Direct" FET fabrication on the crystal surface}

The "direct" fabrication of the single-crystal OFETs, in which a
free-standing OMC is used as the substrate for subsequent
deposition of the contacts and gate dielectric, is not trivial,
because the organic crystals are incompatible with the standard
processes of thin-film deposition/patterning. Fabrication of the
field-effect structures based on single crystals of organic
semiconductors became possible after several innovations have been
introduced both for the source/drain fabrication and for the gate
dielectric deposition \cite{Podzorov03,Podzorov03a}.

\paragraph{Source/drain contacts}

The performance of the organic FETs is often limited by the
injection barriers formed at the interface between the metal
contacts and the semiconductor. The charge injection in such
devices occurs by thermally assisted tunneling of the electrons or
holes through the barrier, whose effective thickness depends on
both gate and source-drain voltages. This is why reducing of the
contact resistance is especially important for proper functioning
of OFETs.

Different routes have been followed for the fabrication of
source/drain contacts on the surface of organic crystals. The
simplest (but also the crudest) is the "manual" application of a
conducting paste. Among the tested materials, the water based
solution of colloidal carbon provided the lowest contact
resistance to organic crystals. A two-component, solvent-free
silver epoxy (Epo-Tek E415G), which hardens at room temperature in
a few hours, has been also used \cite{DeBoer04}. A disadvantage of
this method is that it is difficult to prepare small and
nicely-shaped contacts on hydrophobic OMC surfaces. In addition,
this technique often results in formation of defects (traps) at
the contact/organic interface, as shown, for instance, by the
transport experiments in the space charge limited transport
regime.

The thermal deposition of metals through a shadow mask is a more
versatile method. However, the thermal load on the crystal surface
in the deposition process (mostly because of radiation from the
evaporation boat) has to be painstakingly minimized: deposition
might generate traps at the metal/organic interface, or even
result in the OMC sublimation. The effect of fabrication-induced
traps has been regularly observed in both FET and SCLC
measurements; presence of these traps is also a plausible cause
for irreproducibility of the metal contacts evaporated on top of
organic films used in applied devices \cite{Shen01,Liu01,Rep03}.
This limits the choice of metals to the materials with a
relatively low deposition temperature. As a first (and very crude)
approximation, matching of the metal work-function to the
HOMO(LUMO) levels of OMC for the p-type (n-type) conductivity can
be used as the guideline in the metal selection.

Despite the technological difficulties, successful deposition of
high-quality silver contacts by thermal evaporation has been
recently performed, which demonstrates that the contact
fabrication problems are not intrinsic. In order to minimize
damage of the crystal surface, the authors of Ref.
\cite{Podzorov03a} fixed the deposition rate at a low level ($\sim
1 \ \AA$/s), increased the distance between the evaporation source
and the sample up to 70 cm, maintained the crystal temperature
during the deposition within the range $-20 - 0^{\mathrm{o}}$C by
using a Peltier cooler, and placed a diaphragm near the
evaporation boat to shield IR radiation from the hottest parts of
the boat. It has been also observed that, in order to achieve high
mobilities, it is important to avoid contamination of the channel
surface by metal atoms deposited at oblique angles under the
shadow mask. Such contamination, which dramatically affects the
device performance, presumably occurs because of scattering of
silver atoms from residual gas molecules even at $5 \times
10^{-7}$ Torr, the typical pressure in the chamber for contact
deposition. In order to prevent oblique angle deposition in the
shadowed regions, silver was deposited through a "collimator", a
narrow (4 mm ID) and long (30 mm) tube, positioned close to the
crystal surface. Following this process, high-quality OFETs have
been fabricated on the surface of several organic crystals
(rubrene, TCNQ, pentacene).

In the future, it would be useful to better understand the
mechanism of damaging of organic crystals in the process of
contact fabrication, in order to make the preparation of
high-quality contacts routinely possible with many different
metals. In particular, preparation of high-quality contacts will
help to elucidate the role of the work function of the metallic
electrodes, which seems to play a less prominent role than what
was initially expected (see, e.g., \cite{Gao03}).

\paragraph{Parylene as a novel gate dielectric}

After many unsuccessful attempts, it became clear that sputtering
of Al$_2$O$_3$, as well as other dielectrics, onto the surface of
organic molecular crystals unavoidably results in a very high
density of traps and prohibitively high field-effect threshold:
the field effect is completely suppressed even if the organic
crystals were positioned in the shadow region of the vacuum
chamber, where the deposition rate was zero. Presumably, the OMC
surface is damaged by high-energy particles in the plasma. The
attempts to shield the surface from high-energy charged particles
by electrostatic deflection did not improve the situation. Thermal
deposition of silicon monoxide was also unsuccessful, probably
because of a too high temperature of the deposition source.

The breakthrough in the "direct" fabrication of free-standing
single-crystal OFETs came with using thin polymer films of
parylene as a gate-dielectric material \cite{Podzorov03}. Parylene
coatings are widely used in the packaging applications; the
equipment for parylene deposition is inexpensive and easy to build
(see, e.g., \textit{Parylene Conformal Coatings Specifications and
Properties}, Technical notes, Specialties Coating Systems). This
material with the dielectric constant $\epsilon = 2.65$ forms
transparent pinhole-free conformal coatings with excellent
mechanical and dielectric properties: the breakdown electric field
could be as high as $\sim 10$ MV/cm for the thickness $0.1 \
\mu$m.

\begin{figure}[b]
\centering
\includegraphics[width=8.5cm]{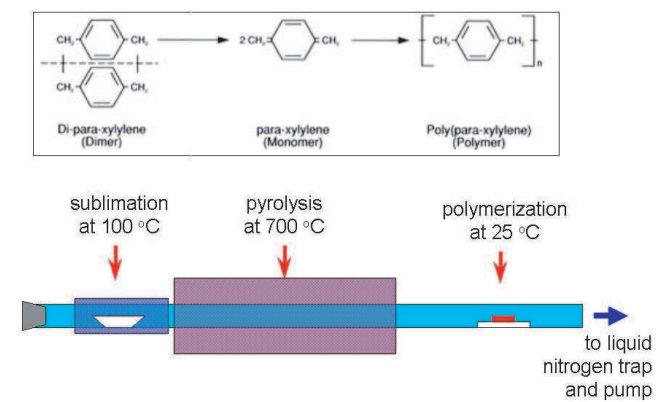}
\caption{Parylene deposition. The top panel shows the reactions
that occur during the deposition: the dimer of parylene sublimes
at $\sim 100^{\mathrm{o}}$C; it splits up to monomers as it enters
the pyrolysis zone at $\sim 700^{\mathrm{o}}$C; the monomers
polymerize in the room temperature zone where the sample is
placed. The bottom panel is a simple schematics of a homemade
deposition chamber. \label{parylenereaction}}
\end{figure}

In Ref. \cite{Podzorov03}, parylene was deposited in a home-made
reactor with three temperature zones (see Fig.
\ref{parylenereaction}). Prior to the deposition, the reactor
(quartz tube) was evacuated to a pressure of $\sim 1$ mTorr. The
dimer \textit{para-xylylene} (generic name, parylene) is vaporized
in the vaporization zone at $\sim 100^{\mathrm{o}}$C, cleaves in
the pyrolysis zone at $\sim 700^{\mathrm{o}}$C, and polymerizes in
the deposition zone (the sample location) at room temperature and
pressure $\sim 0.1$ Torr. The precise value of these parameters
during the parylene deposition is not critical. In Ref.
\cite{Podzorov03}, the parylene deposition rate was $\sim 300 \
\AA$/min for the samples positioned $\sim 35$ cm away from the
pyrolysis zone of the parylene reactor. Parylene was deposited
onto the OMC crystals with pre-fabricated source and drain
contacts with the attached wires (otherwise, contacting the
contact pads might be difficult). The parylene thickness $\sim 0.2
\ \mu$m is sufficient to cover uniformly even the rough
colloidal-graphite contacts. The capacitance of the gate electrode
per unit area, $C_{\mathrm{i}}$, was $C_{\mathrm{i}} = 2 \pm 0.2$
nF/cm$^2$ for a $\sim 1$-$\mu$m-thick parylene film. The output of
working devices with the parylene gate insulator approached $100
\%$ and the parylene films deposited onto organic crystals
withstand multiple thermal cycling between 300 K and 4.2 K.

There are several important advantages of using parylene as the
gate dielectric: (a) it can be deposited while the crystal remain
at room temperature, (b) being chemically inert, it does not react
with OMCs, and (c) the parylene/OMC interface has a low density of
surface states. Apart from that, parylene is a carbon-based
polymer, and its thermal expansion coefficient is likely to be
close to that of most organic crystals (but that remains to be
tested). As it has already been emphasized above, different
thermal expansion/contraction of the crystal and gate dielectric
might result in the stress-induced carrier trapping. In this
regard, the use of parylene is particularly promising for the
operation of OFETs at low temperature. Parylene is also promising
as the gate insulator for the future thin-film, flexible devices,
where flexibility of the gate dielectric is required.

\section{Characteristics of single-crystals OFETs} \label{characterization}

Fabrication of the single-crystal OFETs enables exploration of
\textit{the physical limits on the performance of organic
thin-film FETs}. For the first time, one can study the
characteristics of OFETs not limited by the disorder common for
organic thin films. As the result, many important characteristics
of OFETs, including the charge carrier mobility, the field-effect
threshold, and the sub-threshold slope, have been significantly
improved.

\begin{figure}[t]
\centering
\includegraphics[width=8.5cm]{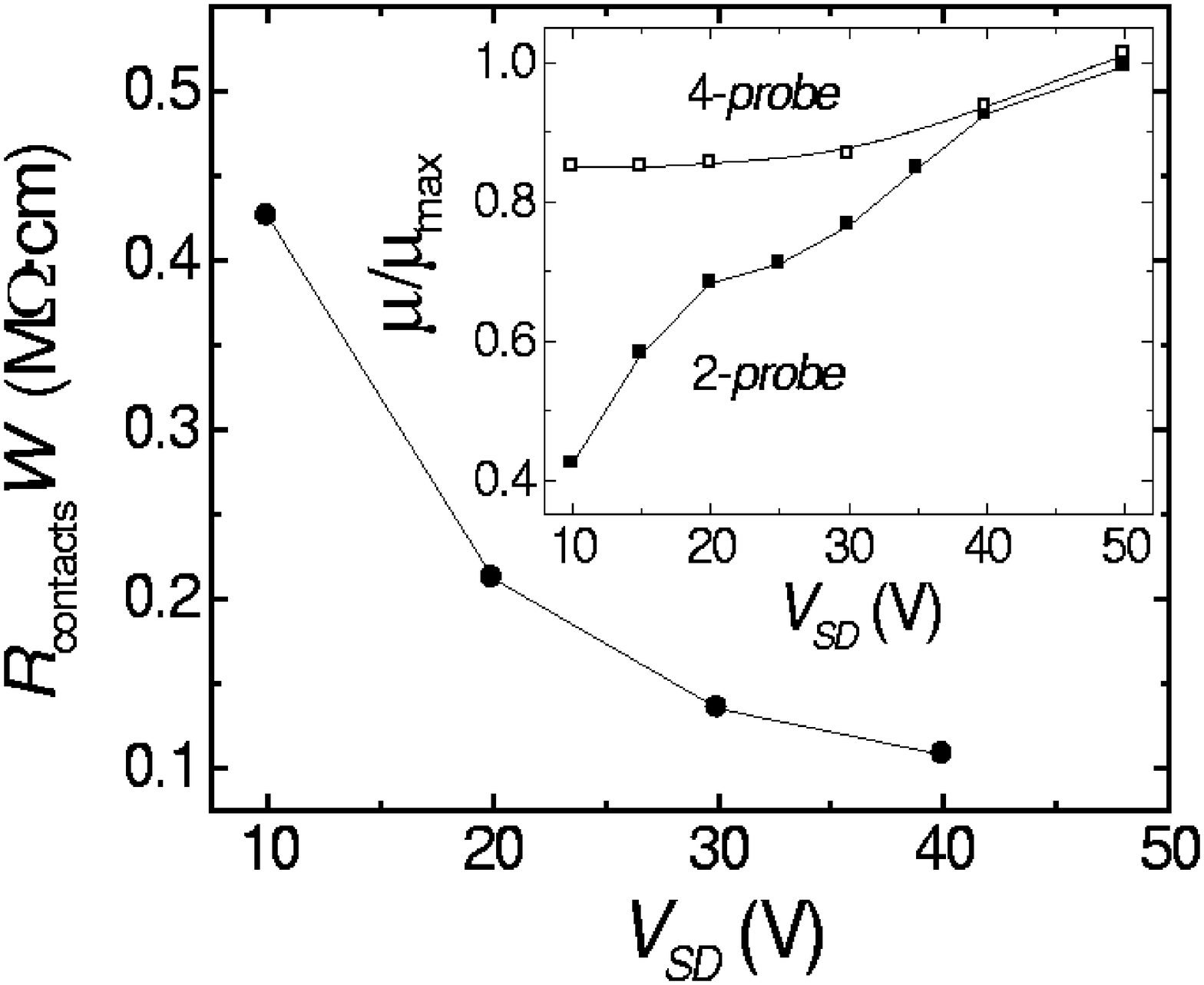}
\caption{The contact resistance normalized by the channel width W
measured using a 4-probe rubrene OFET as a function of the
source-drain voltage ($V_G = -40$ V). The inset: the mobility of
the same device measured in the 4-probe and 2-probe
configurations, normalized by $\mu$ at $V_{SD} = 50$ V.
\label{rubrenefourterm}}
\end{figure}

The organic semiconductors used in OFETs are undoped (or, at
least, not intentionally doped). For this reason, OFETs belong to
the class of injection, or Schottky-limited FETs, in which the
charge carriers are injected into the conduction channel through
the Schottky barriers at the metal/organic interface. For the same
reason, the resistance of source and drain contacts is much higher
than in Si MOSFETs, and depends strongly on the biasing regime.
Since the contact resistance might be comparable or even greater
than the channel resistance (especially at low temperatures), only
the 4-probe measurements provide the intrinsic characteristics of
the conduction channel, not affected by the contact resistance.
However, in the limit of large $V_{G}$ and $V_{SD}$, the contact
resistance becomes small, and the results of 2-probe and 4-probe
measurements typically converge, at least at room temperature (see
Fig. \ref{rubrenefourterm}).

\begin{figure}[b]
\centering
\includegraphics[width=8.5cm]{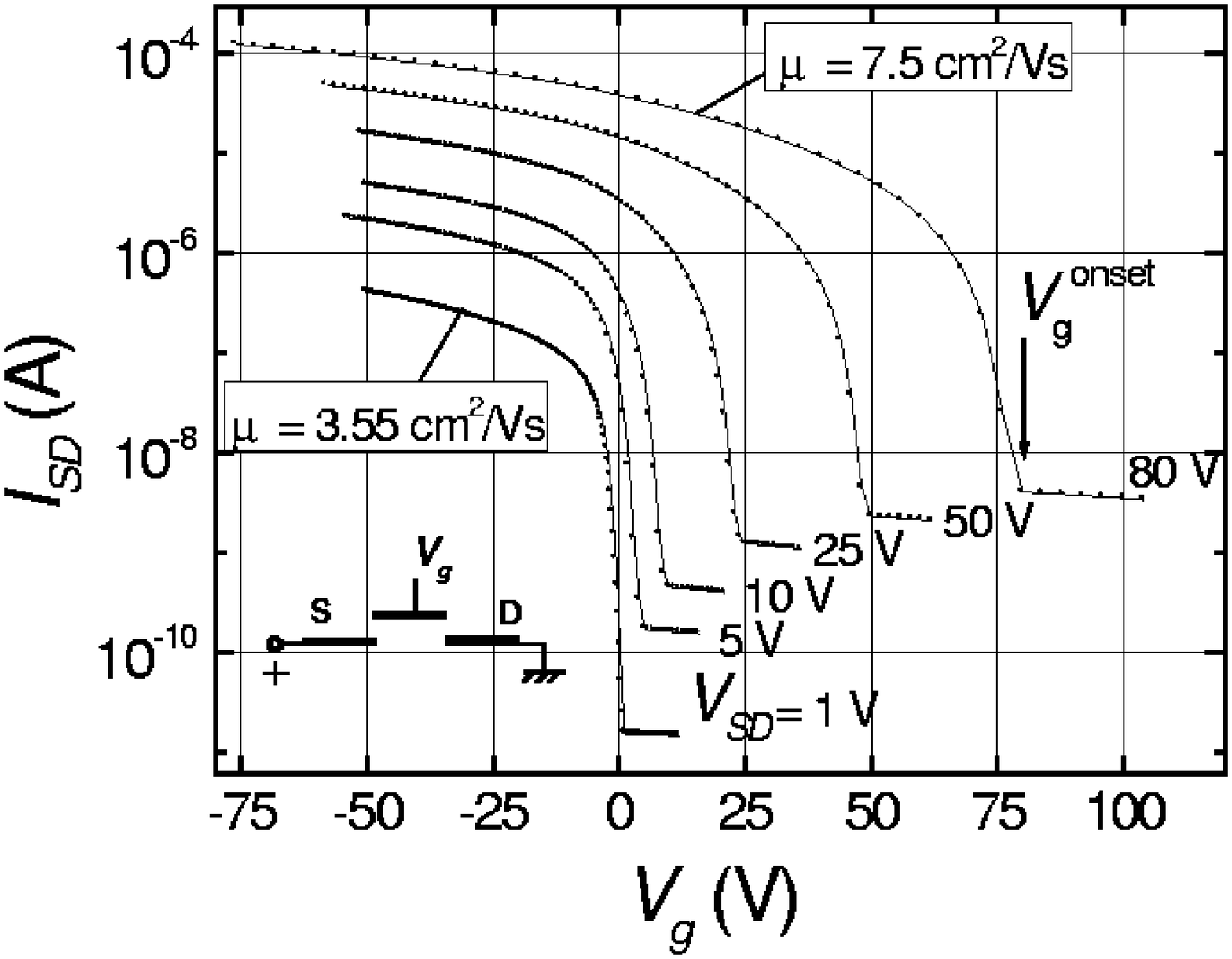}
\caption{The trans-conductance characteristics of an OFET
fabricated on the rubrene single crystal, measured at different
values of the source-drain voltage $V_{SD}$. The in-plane
dimensions of the conducting channel are $L \times W = 1 \times 1$
mm$^2$. \label{Subthresholdrubrene}}
\end{figure}

It is worth mentioning that at this initial stage, when the
research focuses mostly on the study of the intrinsic
field-induced conductivity in organic semiconductors, the biasing
regimes in the experiments with single-crystal OFETs often differ
from the conventional FET biasing \cite{Horowitz98}. The
difference is illustrated in Fig. \ref{Subthresholdrubrene}: the
polarity of the source-drain voltage is chosen to explore a wider
range of the carrier densities. Note also that, because
single-crystal devices have not been optimized for applications
(e.g., the gate insulator is much thicker than in the commercial
devices), the typical values of $V_G$ and $V_{SD}$ are an order of
magnitude greater than that for the conventional Si MOSFETs. The
main characteristics of the single-crystal FETs are summarized
below.

\subsection{Unipolar operation}

\begin{figure}[t]
\centering
\includegraphics[width=8.5cm]{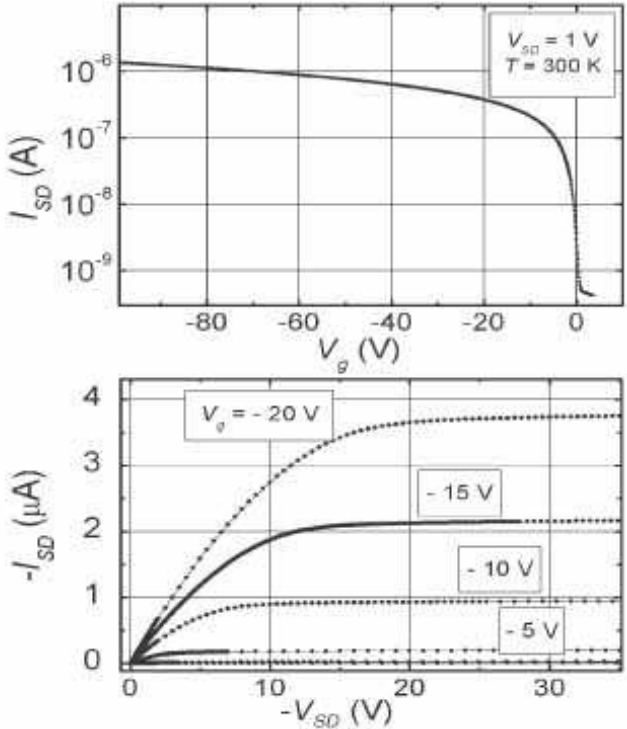}
\caption{Two-probe characteristics of a single-crystal rubrene FET
\cite{Podzorov03}. Upper panel: the dependence of the source-drain
current, $I_{SD}$, on the gate voltage, $V_G$. Lower panel:
$I_{SD}$ versus the bias voltage $V_{SD}$ at several fixed values
of $V_G$. The source-drain distance is 0.5 mm, the width of the
conduction channel is 1 mm, the parylene thickness is $0.2 \
\mu$m. \label{FETcharrubrene}}
\end{figure}

\begin{figure}[t]
\centering
\includegraphics[width=8.5cm]{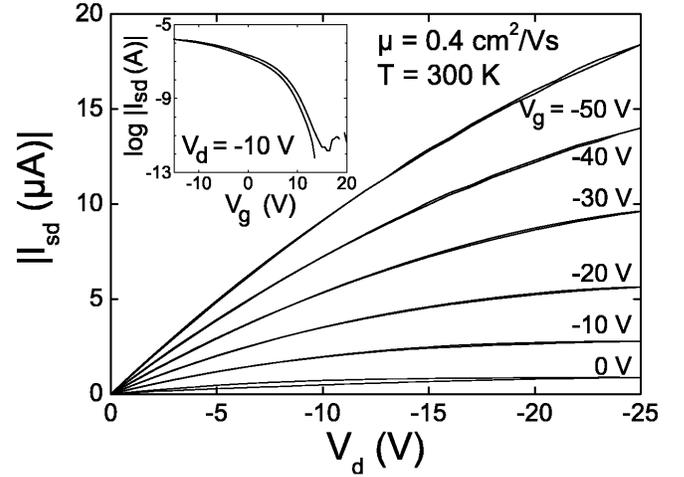}
\caption{Two-probe characteristics of a single-crystal tetracene
FET \cite{DeBoer03}. Source-drain current $I_{SD}$ versus
source-drain voltage $V_{SD}$ measured at different values of
$V_G$. The inset shows the dependence of $\log (I_{SD})$ on $V_G$
at fixed $V_{SD}$, for a different device, which has a mobility
$\mu = 0.05$ cm$^2$/Vs and a threshold voltage $V_{th} \simeq 0.3$
V. From this plot we calculate the subthreshold slope to be 1.6
V/decade. For both devices the source-drain distance is $25 \
\mu$m, the width of the conduction channel is $225 \ \mu$m, and
the SiO$_2$ thickness is $0.2 \ \mu$m. \label{FETchartetracene}}
\end{figure}

All the single-crystal devices fabricated up to date exhibited
unipolar operation. Specifically, the p-type conductivity has been
observed, for instance, in antracene, tetracene, pentacene,
perylene, rubrene, whereas the n-type conductivity has been
observed in TCNQ. Typical transistor characteristics for the
rubrene and tetracene single crystal OFETs are shown in Fig.
\ref{FETcharrubrene} and \ref{FETchartetracene}. In principle, the
unipolar operation can be explained by the choice of metallic
contacts that are efficient injectors of only one type of
carriers. However, the presence of traps that selectively capture
either electrons or holes cannot be excluded. For instance, the
TOF experiments with perylene \cite{Karl99,Karl85a} have shown
that in high quality crystals of this compound, both electrons and
holes are sufficiently mobile at room temperature. However, in the
single-crystal perylene FETs, only the hole conduction has been
observed \cite{DeBoer03a}. One of the reasons for that might be
presence of oxygen in the crystal, which is known to act as a trap
for electrons.

\subsection{Field-effect threshold}

The threshold voltage $V_{\mathrm{th}}$ is a measure of the amount
of charge that it is necessary to induce electrostatically in
order to switch-on electrical conduction in a FET. Using the
equation that describes FET operation
\begin{equation}
I_{SD} = \frac{W}{L} \mu C (V_G - V_{\mathrm{th}}) V_{SD}
\end{equation}
$V_{\mathrm{th}}$ can be obtained by extrapolating the
quasi-linear (high-$V_G$) part of trans-conductance
characteristics $I_{SD}(V_G)$ to zero current (here $W$ and $L$
are the width and the length of the conducting channel,
respectively, and $C$ is the specific capacitance between the
channel and the gate electrode). The charge induced in the
sub-threshold regime fills the traps that immobilize the charge
carriers.

The magnitude of the field-effect threshold voltage depends on
several factors, such as the density of charge traps on the
interface between the organic crystal and the gate dielectric, the
quality of the source/drain contacts (particularly important for
Schottky transistors), and the absence/presence of a "built-in"
conduction channel. Firstly, let's consider the situation when the
built-in channel is absent; this is the case, for example, of the
rubrene devices with parylene gate dielectric
\cite{Podzorov03,Podzorov03a}. The corresponding trans-conductance
characteristics are shown in Fig. \ref{Subthresholdrubrene} on a
semi-log scale. The field-effect onset is observed at a
\textit{positive} gate voltage, similar to the OFET based on
well-ordered pentacene thin-films
\cite{Nelson98,Klauk02,Meijer03}. This behavior resembles the
operation of a "normally-ON" p-type FET with a built-in channel.
The resemblance, however, is superficial: in Ref.
\cite{Podzorov03a}, the sharp onset was always observed at $V_G =
V_{SD}$, which indicates that the channel was induced
electrostatically. Indeed, an application of a positive voltage
$V_{SD}$ to the source electrode in the presence of the gate
electrode $\sim 1 \ \mu$m away from the interface creates a strong
electric field normal to the crystal surface. This field induces
propagation of the conducting channel from the source electrode to
the drain at any $V_G < V_{SD}$. Thus, the single-crystal rubrene
OFETs with parylene gate dielectric are \textit{zero threshold}
devices at room temperature. The zero threshold operation suggests
that the density of the charge traps at the rubrene/parylene
interface is low ($< 10^{9}$ cm$^{-2}$) at room temperature
\cite{notetraps}. However, the situation changes at low
temperatures: the threshold voltage, measured in the 4-probe
configuration, increases with cooling (see Fig.
\ref{rubrenetemp}). This might signal depopulation of the surface
traps that are filled at room temperature. Note that only the
4-probe measurements are essential to study the behavior of
$V_{\mathrm{th}}$; in the 2-probe measurements, an increase of the
non-linear contact resistance with cooling might imitate the
threshold shift.

\begin{figure}[b]
\centering
\includegraphics[width=8.5cm]{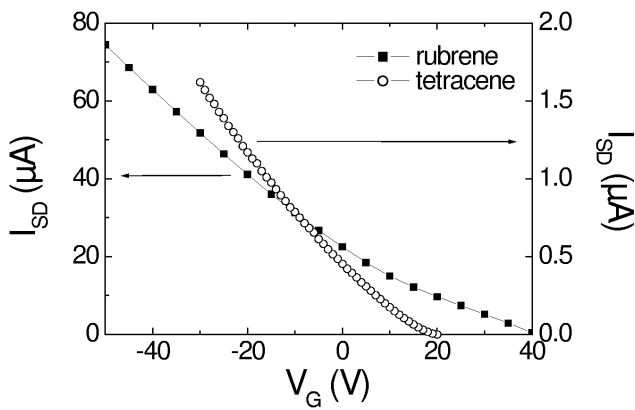}
\caption{Gate-sweeps demonstrating a built-in channel for
electrostatically bonded crystals, both for a tetracene (open
circles) and for a rubrene (filled squares) single-crystal FET.
The $W/L$ ratio is the same for the two devices, $W/L = 0.14$.
\label{builtinchannel}}
\end{figure}

For the OFETs fabricated by the electrostatic bonding of organic
crystals, a relatively large (10 V or more) depletion gate voltage
is often required to completely pinch off the channel (this $V_G$
is positive for the p-type conductivity). This behavior is
illustrated in Fig. \ref{builtinchannel} for rubrene and for
tetracene crystals bonded to the RIE pre-cleaned SiO$_2$ surface
\cite{DeBoer03}, similar behavior has been observed for pentacene
\cite{Takeya03}. The positive threshold has been also observed for
the rubrene crystals bonded electrostatically to the surface of
PDMS rubber stamps \cite{Sundar04}. These observations suggest
that the same microscopic mechanism responsible for electrostatic
bonding might be responsible for inducing the built-in channel on
the organic surface.

\subsection{Sub-threshold slope}

The sharpness of the field-effect onset is characterized by the
sub-threshold slope, $S \equiv \mathrm{d} V_G/\mathrm{d}
(\log{I_{SD}})$. Since this quantity depends on the capacitance of
the insulating layer $C_i$, it is also convenient to introduce the
normalized slope, $S_i \equiv S \cdot C_i$, which permits to
compare more directly the properties of different devices
\cite{Podzorov03a}. For single crystal FETs, even in the devices
with relatively low mobility (the tetracene single-crystal FETs
with $\mu = 0.05$ cm$^2$/Vs \cite{DeBoer03}), the observed
normalized sub-threshold slope $S_i = 28$
V$\cdot$nF/decade$\cdot$cm$^2$ was comparable with that for the
best pentacene TFTs ($S_i = 15 - 80$
V$\cdot$nF/decade$\cdot$cm$^2$
\cite{Dimitrakopoulos99,Lin97,Dimitrakopoulos99a}). The
high-mobility single-crystal rubrene OFETs with $\mu \simeq 5 - 8$
cm$^2$/Vs exhibit a sub-threshold slope as small as $S = 0.85$
V/decade, which corresponds to $S_i = 1.7$
V$\cdot$nF/decade$\cdot$cm$^2$ \cite{Podzorov03a}. This value is
an order of magnitude better than what has been achieved in the
best organic TFTs; it also compares favorably with $\alpha$-Si:H
FETs, for which $S_i \simeq 10$ V$\cdot$nF/decade$\cdot$cm$^2$ has
been reported \cite{Kanicki91}.

\begin{figure}[t]
\centering
\includegraphics[width=8.5cm]{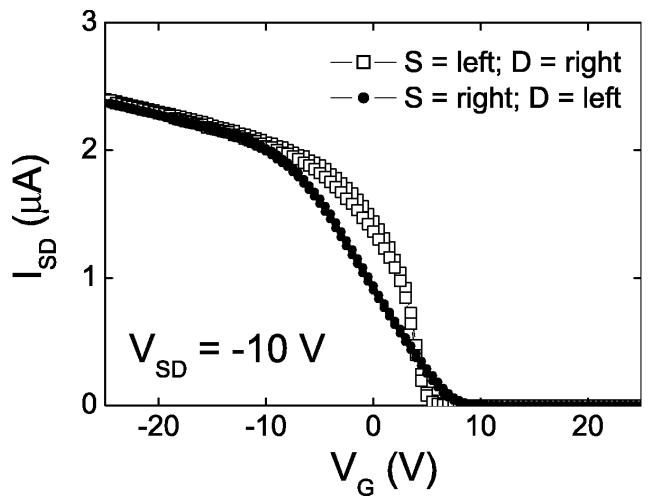}
\caption{$V_G$-sweeps of a rubrene FET fabricated by electrostatic
bonding. The two different curves are obtained by interchanging
the source and the drain, while the source-drain voltage is the
same (-10 V) in both cases. The influence of the contacts is
visible in the sub-threshold region of the $V_G$-sweeps, which
shows a clearly asymmetric behavior, in spite of the long channel
length ($1200 \ \mu$m). At higher values of $V_G$, the electrical
characteristics are independent of the source/drain configuration.
The channel width of this device is approximately $200 \ \mu$m.
\label{posnegpolarity}}
\end{figure}

It is commonly believed that the sub-threshold slope is mainly
determined by the quality of insulator/semiconductor interface
\cite{Sze81}. This is definitely the case for Si MOSFETs, where
the resistance of source and drain contacts is low and does not
depend on the gate voltage. In contrast, the contact resistance in
the Schottky-type OFETs is high, it depends non-linearly on $V_G$
- as the result, the sub-threshold slope might reflect the quality
of contacts rather than the insulator/semiconductor interface. The
effect of the contacts is illustrated in Fig. \ref{posnegpolarity}
for a rubrene device fabricated by electrostatic bonding. In this
device, interchanging the source and drain contacts results in
different sub-threshold $V_G$ characteristics. At higher values of
$V_G$, when the conducting channel is formed, the electrical
characteristics are symmetric, i.e. they are not sensitive to the
source/drain configuration. Note that in this device the contacts
dominate the behavior in the sub-threshold region even though the
length of the channel is considerable (1.2 mm; the channel width
is approximately 0.2 mm).

\subsection{Double-gated rubrene FETs}

\begin{figure}[b]
\centering
\includegraphics[width=8.5cm]{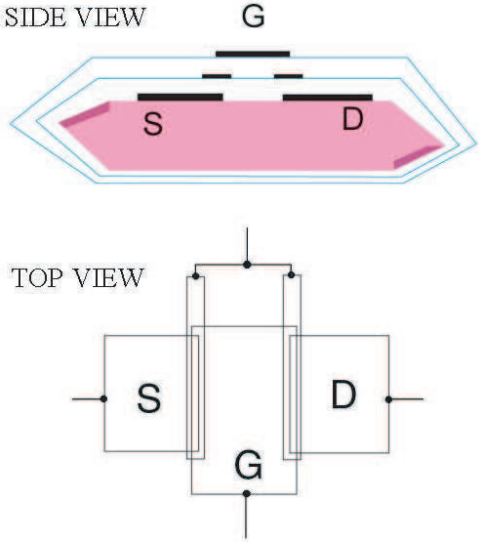}
\caption{Schematic design of the double-gated FET. Additional to
the conventional source, drain and gate-electrode, also a
complimentary gate electrode is deposited, which is separated from
the other electrodes by parylene layers. The complementary gate
overlaps with the source and drain contacts, and it serves to
control the contact resistance. \label{picturedoublegate}}
\end{figure}

The conventional method for the fabrication of low-resistance
contacts in Si MOSFETs is based on ion implantation of dopants
beneath the contact area. Unfortunately, a similar technique has
not been developed for OFETs yet. In this situation, the contact
resistance can be reduced by using the so-called double gate, a
trick that has been successfully applied for the study of
high-mobility Si MOSFETs at low temperatures \cite{Klapwijk98}.
Schematic design of such device, fabricated at Rutgers, is shown
in Fig. \ref{picturedoublegate}.

Two separately biased gate electrodes are deposited on the surface
of a rubrene crystal with pre-formed source and drain contacts:
the main gate and the complementary gate. These electrodes are
isolated from the crystal and each other by a layer of parylene
($\sim 1 \ \mu$m thick). The complimentary gate electrode (closest
to the surface) consists of two stripes, connected together, that
overlap with the source and drain contacts. The resistance of
Schottky barriers and the charge density in the channel of this
device can be controlled separately by applying different voltages
to the complimentary gate, $V_{Gc}$, and to the main gate
electrode, $V_G$.

\begin{figure}[t]
\centering
\includegraphics[width=8.5cm]{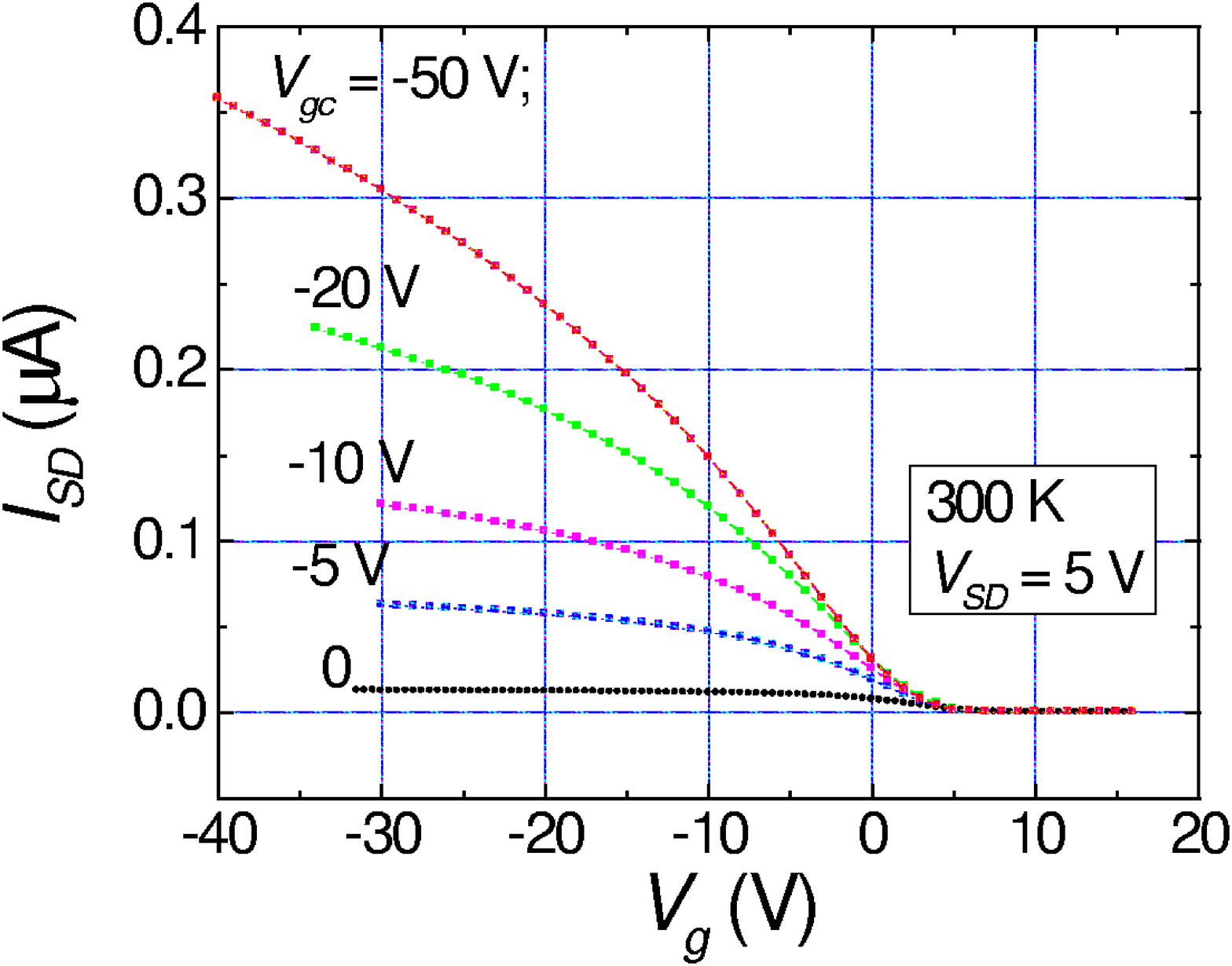}
\caption{The trans-conductance characteristics of the double-gated
rubrene FET. A set of $I_{SD}(V_G)$ curves is shown for a fixed
$V_{SD} = 5$ V and several negative complimentary gate voltages
$V_{Gc}$. \label{resultsdoublegate}}
\end{figure}

The trans-conductance characteristics of the double-gated rubrene
FET, $I_{SD}(V_G)$, are shown in the Fig. \ref{resultsdoublegate}
for several values of $V_{Gc}$ and a fixed $V_{SD} = 5$ V. Large
negative voltage $V_{Gc}$ greatly reduces the contact resistance,
and the regime of low carrier densities becomes easily accessible.
However, the price for this is the non-linearity of
trans-conductance characteristics at $V_G > V_{Gc}$: a portion of
the semiconductor surface beneath the complimentary gate electrode
is screened from the field of the main gate electrode, and its
resistance does not depend on $V_G$.

\subsection{Mobility}

The mobility of carriers at the surface of organic crystals can be
estimated from the linear portion of the trans-conductance
characteristics, where the conductivity of the channel, $\sigma =
e n \mu$, varies linearly with the density of mobile field-induced
charges, $n$. The "intrinsic" mobility, not limited by the contact
resistance, can be estimated from the 4-probe measurements as
\cite{Sze81}
\begin{equation} \label{mobfourterm}
\mu = (\frac{L}{W C_i V})(\frac{dI_{SD}}{dV_G})
\end{equation}
where $V$ is the potential difference between the voltage probes
at a distance $L$ from each other, $W$ is the channel width, and
$C_i$ is the specific capacitance between the gate electrode and
the conduction channel. For the 2-probe measurements, $L$ in Eq.
\ref{mobfourterm} corresponds to the total length of the
conduction channel, and $V$ to the source-drain voltage $V_{SD}$.
The latter measurements usually provide a lower estimate for
$\mu$, which approaches the intrinsic $\mu$ value with increasing
$V_G$ and $V_{SD}$ (see Fig. \ref{rubrenefourterm}). Eq.
\ref{mobfourterm} is based on the assumption that all charge
carriers induced by the transverse electric field above the
threshold are mobile, and their density is given by:
\begin{equation} \label{chargedensity}
n = \frac{C_i (V_G - V_{\mathrm{th}})}{e}
\end{equation}
This assumption has not been fully justified yet. For comparison,
in a different type of FETs with comparable values of $\mu$,
amorphous silicon ($\alpha$-Si:H) FETs, this is not the case:
above the threshold, most of the induced charge in these devices
goes into the "tail" (localized) states with only a small fraction
going into the conduction band (see, e.g., \cite{Shur89}). The
latter model of multiple thermal trapping and release of carriers
involving shallow traps is not appropriate for OFETs, where charge
transport cannot be described in terms of band transport owing to
polaronic effects \cite{Silinsh94,Wu97}. Some justification of
estimate \ref{chargedensity} is provided by observations of
$V_G$-independent mobility and the mobility increase with cooling,
in a sharp contrast with the behavior of the $\alpha$-Si:H FETs.
Note that, contrary to the conventional inorganic FETs, the
density of polaronic charge carriers in OFETs cannot be estimated
from the Hall-type experiments, at least at high temperatures
where hopping processes govern the charge transport (for
discussion of the Hall effect in the polaronic hopping regime, see
Ref. \cite{Friedman78}).

\begin{figure}[b]
\centering
\includegraphics[width=8.5cm]{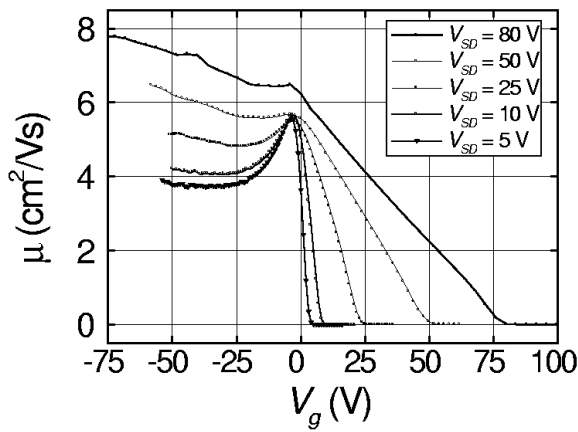}
\caption{The dependencies $\mu(V_G)$ for the single-crystal
rubrene OFETs, calculated from 2-probe measurements. The peak of
$\mu$ near the zero $V_G$ are the artifact of the 2-probe
measurements; it is related to a rapid decrease of the total
resistance of the source and drain contacts with increasing charge
density. Note that the source is positive with respect to the
grounded drain, so that for negative gate voltage the transistor
is not in the saturation regime. \label{mobilityrubrene}}
\end{figure}

The room temperature mobility of the field-induced carriers varies
over a wide range for different organic single crystals. The
following values of $\mu$ have been reported: tetracene - 0.4
cm$^2$/Vs \cite{DeBoer03}, pentacene - 0.5 cm$^2$/Vs
\cite{Takeya03,Stassen04}, rubrene - 10 cm$^2$/Vs (up to 15
cm$^2$/Vs in recent unpublished measurements \cite{Sundar04}),
TCNQ - 1 cm$^2$/Vs \cite{Podzorov04}). For most of these
materials, comparable values of $\mu$ have been obtained for FETs
fabricated by both techniques of electrostatic bonding and direct
fabrication on the crystal surface - another indication of the
fact that, in many cases, the measurements with single-crystal
OFETs probe the electronic properties of the crystals, at least at
room temperature, and are not affected by artifacts due to the
device fabrication.

The $\mu$ values for single-crystal devices are comparable or
greater than the corresponding values of $\mu$ reported for the
best thin-film devices (see, e.g., \cite{DeBoer03,Gundlach02}). To
our knowledge the only exception is pentacene, for which the
highest measured TFT mobility is 3 cm$^2$/Vs. Recently however, a
room temperature mobility estimated by in-plane SCLC measurements
as high as $\sim 30$ cm$^2$/Vs has been reported
\cite{Jurchescu04}. Work on the fabrication of FETs based on these
pentacene single-crystals is in progress \cite{Stassen04a}.

Two inter-related factors play an important role in the mobility
improvement in single-crystal OFETs with respect to the organic
TFTs. Firstly, the single crystal surfaces are free from the
inter-grain boundaries that might limit significantly the mobility
in the thin-film devices \cite{Horowitz03a}. Secondly, the
experiments with organic single crystals demonstrated for the
first time that the mobility might be strongly anisotropic (see
Sec. \ref{anisotropy}). Thus, in the experiments with single
crystals, there is a possibility to choose the direction of the
maximum mobility. In the case of elongated, needle-like crystals
(rubrene, pentacene, etc.), the direction of maximum mobility
coincides with the direction of the fastest crystal growth with
the strongest inter-molecule interactions. At the same time, the
mobility of carriers in OTFTs with the grains oriented randomly
with respect to the current is an "angle-averaged" quantity: the
grains oriented along the axis of the minimum mobility will have a
much higher resistance.

The mobility in the single-crystal OFETs depends much less on the
carrier density and the source-drain voltage than that in the
organic TFTs. In the latter case, a pronounced increase of $\mu$
with $V_G$ is observed due to the presence of structural defects
\cite{Dimitrakopoulos02}; because of the strong $\mu(V_G)$
dependence in the TFTs, a large $V_G \geq 100$ V (for a typical
$0.2 \ \mu$m thick SiO$_2$ gate insulator) is often required to
realize higher mobilities, comparable to that of $\alpha$-Si:H
FETs ($\sim 0.5$ cm$^2$/Vs). The typical dependence of the
"2-probe" $\mu$ on the gate voltage for rubrene OFETs is shown in
Fig. \ref{mobilityrubrene}. The maximum of $\mu(V_G)$ near the
zero $V_G$ and an apparent increase of the mobility with $V_{SD}$
are the artifacts of the 2-probe measurements; these artifacts are
caused by a strong $V_G$-dependence of the resistances of the
source and drain contacts \cite{Chwang00}. At sufficiently large
negative gate voltage ($V_G \leq -20$ V), $\mu$ becomes almost
$V_G$-independent (the variations ${\Delta \mu}/\mu$ do not exceed
$15 \%$). The dependence of $\mu$ on $V_{SD}$, measured for the
rubrene OFETs in the 2- and 4-probe configurations are compared in
the inset to Fig. \ref{rubrenefourterm}. The 4-probe data reflect
the "intrinsic" charge carrier mobility, which is only weakly
dependent on $V_{SD}$. The 2-probe data converge with the 4-probe
data at high $V_{SD}$ owing to a lower contact resistance. Similar
independence of the mobility of $V_G$ has been observed for
tetracene single crystal FETs \cite{DeBoer03}.

\begin{figure}[t]
\centering
\includegraphics[width=8.5cm]{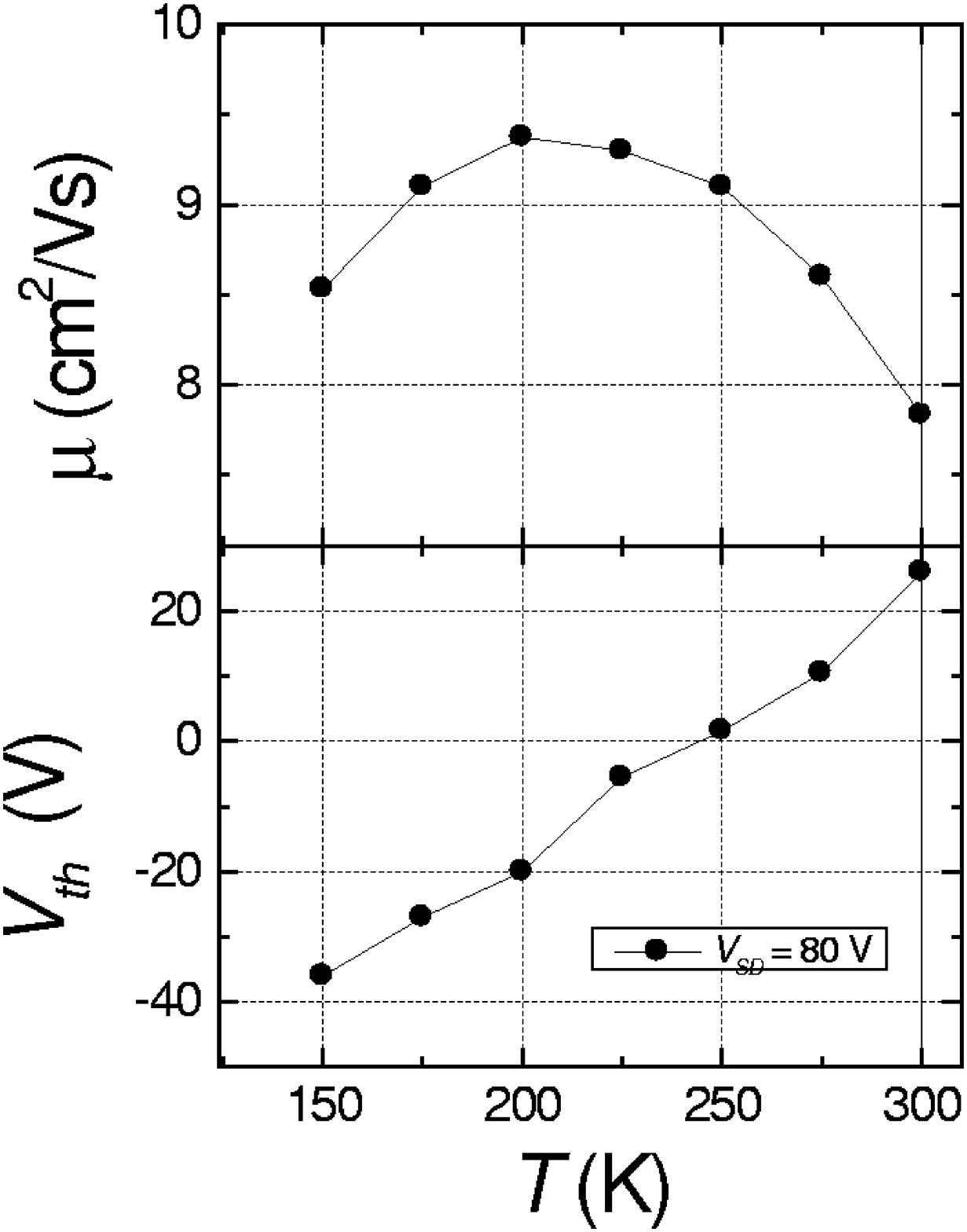}
\caption{Temperature dependence of a rubrene FET with a room
temperature mobility of 7.5 cm$^2$/Vs. These measurements are
performed in 4-probe configuration, at large positive source-drain
voltage, $V_{SD} = 80$ V. Top panel: The mobility shows a
non-monotonic behavior as a function of temperature, with an
optimum value around $\sim 200$ K. Bottom panel: With lowering
temperature the threshold voltage becomes smaller and eventually
even changes sign. \label{rubrenetemp}}
\end{figure}

\begin{figure}[t]
\centering
\includegraphics[width=8.5cm]{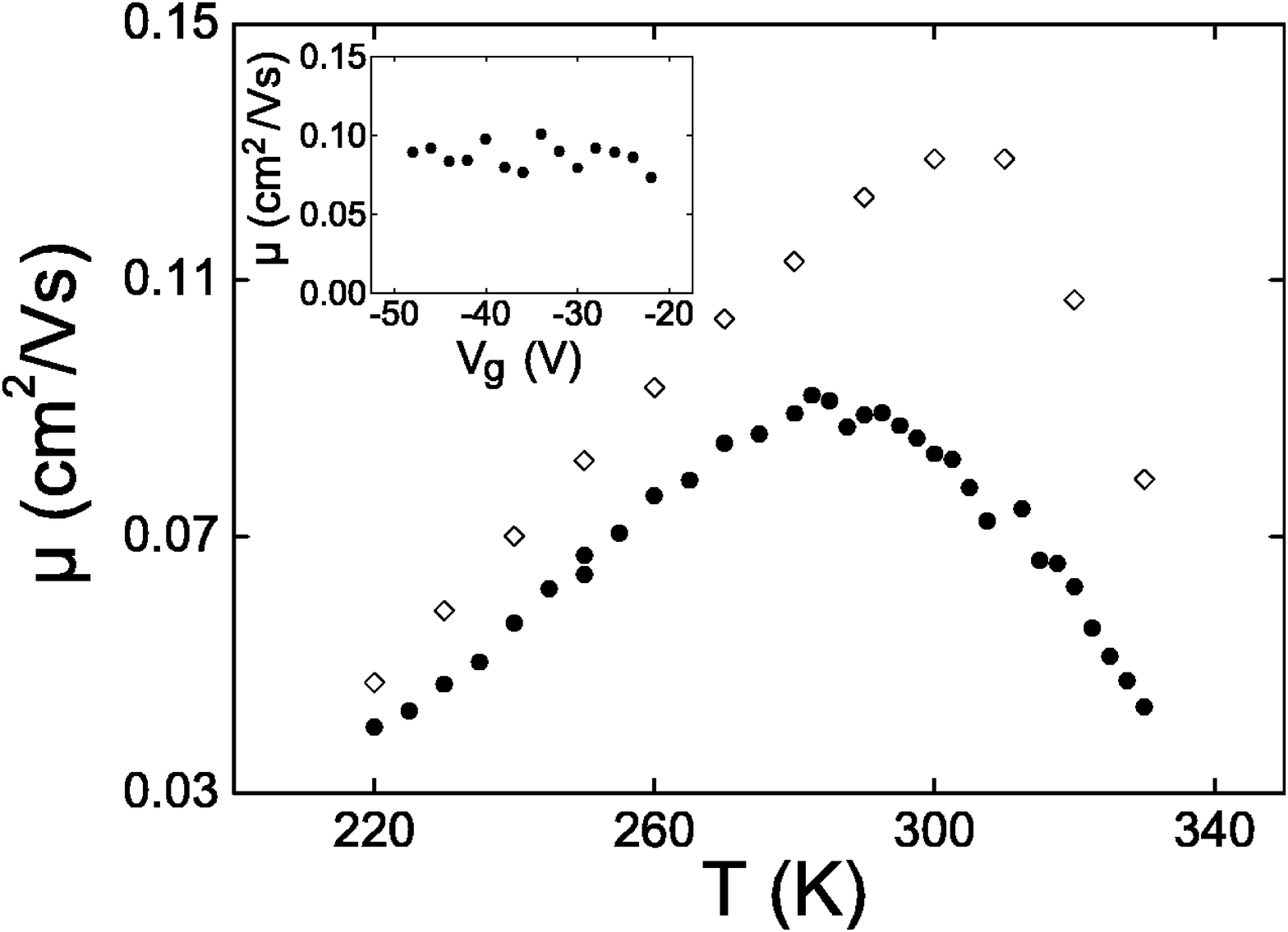}
\caption{Temperature dependence of $\mu$ for two different
tetracene FETs, measured at large negative gate voltage. The inset
illustrates that at large negative gate voltage, -20 to -50 V,
where the highest mobility is observed, $\mu$ is essentially
independent of $V_G$. \label{tetracenetemp}}
\end{figure}

A non-monotonous temperature dependence of $\mu$ has been observed
on the devices with highest (for a given material) mobilities:
with cooling from room temperature, the mobility initially
increases and then drops sharply below $\sim 100$ K. The
temperature dependencies of the mobility for the rubrene (4-probe
measurements) and tetracene (2-probe measurements) FETs are shown
in Fig. \ref{rubrenetemp} and \ref{tetracenetemp}. Similar trend
has been observed for the high-mobility pentacene OFETs
\cite{Nelson98,Takeya03}. There is no correlation between the
absolute value of $\mu$(300 K) and its temperature dependence:
similar dependencies $\mu(T)$ have been observed for the tetracene
OFETs with $\mu \simeq 0.1$ cm$^2$/Vs and rubrene OFETs with $\mu
> 10$ cm$^2$/Vs. The low-temperature drop
of $\mu$ can be fitted by an exponential dependence $\mu = \mu_0
\exp(-T/{T_0})$ with the activation energy $T_0 \simeq 50-150$
meV. Observation of this drop in the 4-probe measurements
indicates that the exponential decrease of the mobility at low
temperatures is not an artifact of the 2-probe geometry and
rapidly increasing contact resistance.

There are some indirect indications that the room-temperature
mobility in the devices with a low density of defects (high-purity
crystals and high quality of organic surface) approaches its
intrinsic value. The increase of $\mu$ with cooling, which is
usually considered as a signature of the intrinsic transport,
correlates with observation of the mobility anisotropy (see Sec.
\ref{anisotropy}). The mobility drop, observed with further
cooling, resembles the data obtained in the TOF experiments for
not-so-pure crystals - this drop is likely caused by trapping of
carriers by shallow traps, which can be active above the
field-effect threshold due to thermal excitations in the system.
An increase of the threshold voltage with cooling, clearly seen in
Fig. \ref{rubrenetemp}, indicates that the trap concentration is
relatively large even in the best crystals that have been used so
far for the OFET fabrication. At present, the quantitative
description of the polaronic transport, in general, and
interaction of polarons with shallow traps in the FET experiments,
in particular, is lacking. More systematic four-probe measurements
of $\mu$ in different molecular crystals, along with the
theoretical efforts on description of polaronic transport in
systems with disorder, are needed to understand the origin of the
observed temperature dependence of the mobility.

\subsection{Mobility anisotropy on the surface of organic crystals.} \label{anisotropy}

Because of a low symmetry of organic crystals, one expects a
strong anisotropy of their transport properties. Indeed, a strong
anisotropy of the polaronic mobility with respect to the
crystallographic orientation has been demonstrated by the TOF
experiments \cite{Karl99,Karl85a,Karl01}. Observation of the
mobility anisotropy can be considered as a prerequisite for
observation of the intrinsic (not limited by disorder) transport
in organic semiconductors. The measurements on thin-film organic
transistors never revealed such anisotropy. By eliminating grain
boundaries and other types of defects, fabrication of the
single-crystal OFETs allow for the first time to address the
correlation between the molecular packing and the anisotropy of
the charge transport on the surface of organic crystals.

\begin{figure}[t]
\centering
\includegraphics[width=8.5cm]{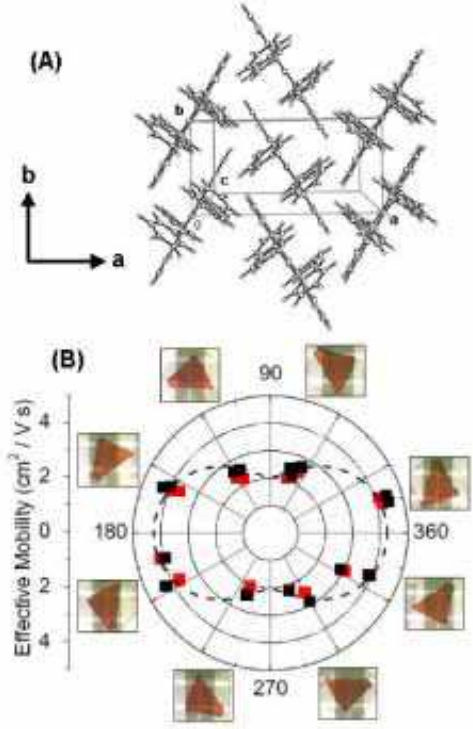}
\caption{(a) Molecular packing in the \textit{a-b} plane of the
rubrene crystal. The direction of maximum mobility corresponds to
the \textit{b}-axis. (b) The angular dependence of the mobility
for a rubrene crystal, measured by a 2-probe rubber-stamp device
at room temperature. Black and red dots correspond to the $\mu$
values extracted from the linear and saturation regime of the FET
$I-V$ curves. The experiment was performed twice to ensure the
reproducibility of this result. \label{angulardep}}
\end{figure}

To probe the mobility anisotropy on the \textit{a-b} surface of
the rubrene crystals, the researchers in Ref. \cite{Sundar04}
exploited an advantage of the PDMS stamp technique, which allows
re-establishing the contact between the stamp and some organic
crystals without damaging the surface. In this experiment, the
crystal was rotated in a step-wise fashion, after each rotation
the 2-probe stamp was re-applied, and the mobility was measured in
both linear and saturation regime to exclude the effect of
contacts. Figure \ref{angulardep}b shows the data for two
$360^{\mathrm{o}}$ rotations, to demonstrate reproducibility of
the results. Black and red symbols correspond to the $\mu$ values
extracted from the linear and saturation regimes of the FET $I-V$
characteristics, respectively. The agreement between these values
confirms that the contact effects can be neglected. Combination of
the orientation-dependent field-effect transport measurements with
the Laue x-ray analysis of the crystals shows that the direction
of the highest mobility coincides with the \textit{b}-axis (see
Fig. \ref{angulardep}a).

\begin{figure}[t]
\centering
\includegraphics[width=8.5cm]{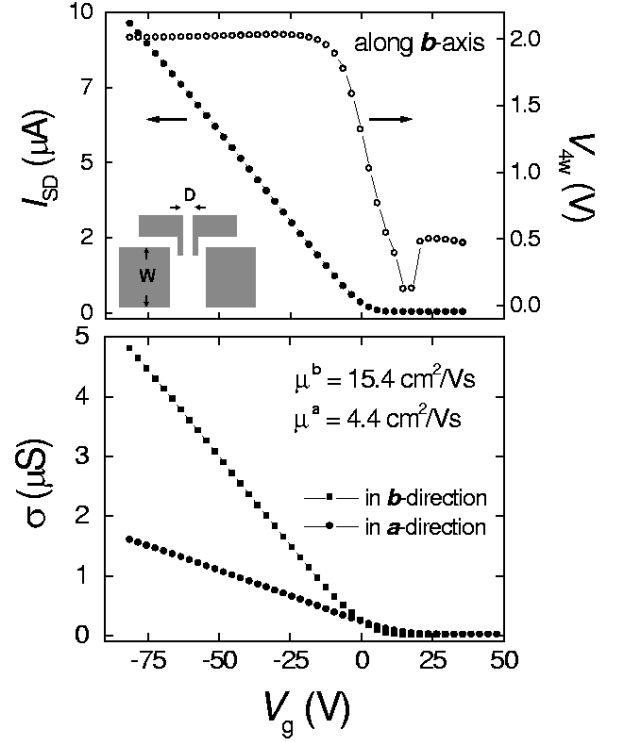}
\caption{Top panel: Four-probe measurements of the charge
transport along the \textit{b}-axis of rubrene crystal: the
source-drain current $I_{SD}$ (closed circles) and the voltage
difference between the voltage contacts $V_{4w}$ (open circles)
are shown as a function of the gate voltage $V_G$. The inset
illustrates the contact geometry. Lower panel: The channel
conductivity $\sigma$ as a function of $V_G$ extracted from the
4-probe measurements along two crystallographic axis.
\label{aandbaxis}}
\end{figure}

The channel conductivity, $\sigma(V_G) \equiv I_{SD}/V_{4w}$,
extracted from the 4-probe measurements along the directions of
maximum and minimum mobility (the \textit{b} and \textit{a}
crystallographic axes, respectively) are shown in Fig.
\ref{aandbaxis}. The mobility $\mu = (D/W C_i)(\mathrm{d}
\sigma/\mathrm{d} V_G)$ along the \textit{b}-axis is almost 4
times greater than the mobility along the a axis.

\subsection{Preliminary results for the OFETs with high-\textit{k} dielectrics.}

\begin{figure}[t]
\centering
\includegraphics[width=8.5cm]{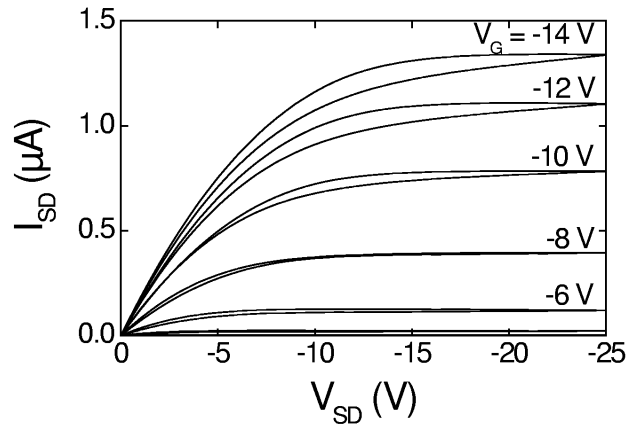}
\caption{Two-probe characteristics of a single-crystal tetracene
FET with Ta$_2$O$_5$ as a dielectric. The figure shows the
source-drain current $I_{SD}$ versus source-drain voltage $V_{SD}$
measured at different values of $V_G$. The source-drain distance
is $25 \ \mu$m, the width of the conduction channel is $225 \
\mu$m, and the Ta$_2$O$_5$ thickness is $0.35 \ \mu$m. The
mobility of the device is 0.03 cm$^2$/Vs. \label{FETcharTaO}}
\end{figure}

The characteristics of OFETs with high-\textit{k} gate dielectrics
(see Sec. \ref{adhesion}) are in many respects similar to those
for devices with the SiO$_2$ gate dielectric. The $I-V$ curves
measured for a tetracene FET electrostatically bonded onto a
Ta$_2$O$_5$ insulating layer are shown in Fig. \ref{FETcharTaO}.
Although the leakage gate current is normally higher than that for
SiO$_2$, it is still significantly smaller than the source-drain
current. The maximum charge density that has been reached so far
is $\sim 5 \times 10^{13}$ cm$^{-2}$.

Despite apparent similarity, there is a significant difference
between FETs fabricated on SiO$_2$ and on Ta$_2$O$_5$: the
characteristics of high-\textit{k} OFETs degrade substantially
upon successive measurements. Specifically, the source-drain
current measured at a fixed source-drain voltage is systematically
decreased when subsequent scans of the gate voltage are performed.
In general, the higher the applied gate voltage, the stronger the
degradation. At the same time, the threshold voltage
systematically increases to higher (negative) values.

\begin{figure}[t]
\centering
\includegraphics[width=8.5cm]{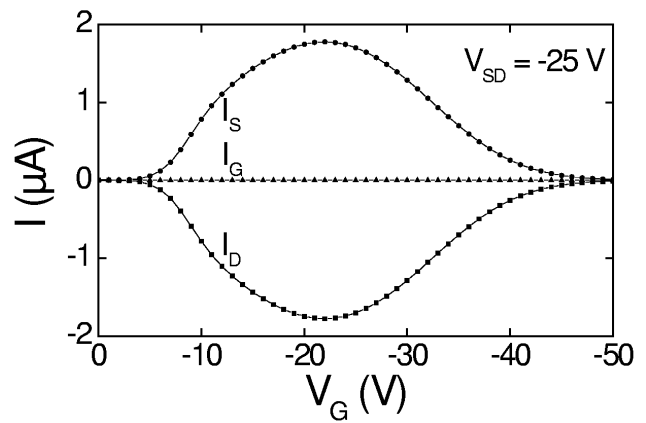}
\caption{Two-probe gate-sweep of the same tetracene device as in
figure \ref{FETcharTaO}. The source current $I_S$ and the drain
current $I_D$ have the same magnitude and opposite sign, as
expected. The leakage current to the gate, $I_G$, is typically
much smaller than the source-drain current. Ramping the gate
voltage up to high value results in a non-monotonous dependence of
the source-drain current. This is due to degradation of the device
caused by the small, but finite, current that is leaking through
the gate insulator. The degradation is irreversible: when the same
scan is repeated again, there is no field-effect and the current
through the device is zero. \label{gatesweepTaO}}
\end{figure}

For the case of Ta$_2$O$_5$, this degradation eventually results
in a complete suppression of the field effect. This is shown in
Fig. \ref{gatesweepTaO} where the source, the drain and the
leakage current are plotted as a function of the gate voltage, for
a fixed source/drain voltage ($V_{SD} = -25$ V). Instead of
increasing linearly with $V_G$ as in conventional FETs, the
source/drain current exhibits a maximum (at $V_G \simeq -22$ V in
this device) and then decreases to zero upon further increasing
$V_G$. This degradation is completely irreversible: no measurable
source/drain current was detected when $V_G$ was decreased to 0
and swept again to high voltage.

The experiments show that current leaking from the gate is the
cause of the FET degradation. This current, typically much smaller
than the source-drain current (see Fig. \ref{gatesweepTaO}), has a
different effect: the electrons leaking through the gate
insulator, accelerated in a strong electric field due to the
voltage applied to the gate electrode, damage the organic
crystals. The precise microscopic mechanism responsible for the
damage is still unclear, but it is likely that the injection of
high-energy electrons introduces surface defects (locally
disrupting the crystal due to the accumulation of negative charge)
and result in formation of traps (e.g., "broken" individual
moleucles). This would account for the observed decrease in
mobility and shift in threshold voltage.

Optimization of the growth of high-\textit{k} dielectrics in a
near future is possible, since it is known how to reduce the
leakage current by several orders of magnitude (by sputtering the
high-\textit{k} materials onto a heated substrate
\cite{Fleming00}). This is known to bring it to the level observed
in SiO$_2$ ($\leq 10^{-10}$ A/cm$^2$ for a SiO$_2$-thickness of
200 nm), at which the irreversible damage introduced in the
crystal is minimal. This improvement will allow stable FET
operation at a charge density $\sim 1 \times 10^{14}$ cm$^{-2}$.

\section{Conclusion} \label{conclusion}

The development of single-crystal OFETs offers a great opportunity
to understand better the charge transport in organic
semiconductors and to establish the physical limits on the
performance of organic field-effect transistors. Below we briefly
summarize the recent progress in this field, and attempt to
formulate the research directions that, from our viewpoint, need
to be addressed in the near future.

One of the central issues in the field-effect experiments with
organic semiconductors is realization of the \textit{intrinsic
(not limited by disorder)} charge transport on the organic
surface. In disorder-free organic semiconductors, several regimes
of polaronic transport are expected (see, e.g.,
\cite{Silinsh94,Fratini03}), between the limiting cases of the
coherent motion in extended states (band-like polaronic transport)
at low temperatures and the incoherent hopping at high
temperatures. All these regimes are intrinsic in the
aforementioned sense and all of them require better understanding:
the incoherent polaronic hopping is pertinent, for instance, to
the devices functioning at room temperature, whereas the
realization of band-like polaronic transport is the Holy Grail of
basic research.

We have witnessed significant progress towards realization of this
goal. In this respect, an important characteristic of the first
generation of single-crystal OFETs, whose significance should not
be underestimated, is the device reproducibility and consistency
of experimental observations - e.g., similar mobilities have been
observed for the single-crystal OFETs based on the same organic
compound but fabricated by different teams using different
fabrication techniques. Although it might be taken for granted
that this should be the case, such reproducibility has never been
achieved in thin-film organic field effect transistors: the
organic TFTs are notoriously known for a large spread of
parameters, even if they were prepared under nominally identical
conditions \cite{Dimitrakopoulos02,Campbell01}. This
irreproducibility is still a problem despite the fact that OTFTs
have been the subject of intense research for a much more extended
period of time than single crystal OFETs.

The reproducibility of the OFET parameters is a necessary
condition for investigation of the intrinsic electronic properties
of organic semiconductors; the reproducibility, however, does not
guarantee that the electrical behavior of the first-generation
single crystal OFETs reflects the \textit{intrinsic} electronic
properties of organic molecular materials. Indeed, all OFETs
investigated by different research groups have been fabricated on
the basis of crystals grown under similar conditions, and,
presumably, with a similar concentration of defects. Thus, the
question remains: how close are we to the realization of the
intrinsic transport in organic single crystal FETs?

For different temperature ranges, the answer to this central
question will be different. There are many indications that at
room temperature, the best single-crystal OFETs demonstrate the
intrinsic behavior, and the room-temperature mobility of these
devices approaches its intrinsic value. For example, observation
of the anisotropy of the mobility of field-induced charges in the
\textit{a-b} plane \cite{Sundar04} reveals for the first time (in
the FET experiments) correlation between the charge transport and
the underlying lattice structure. Another important observation is
a moderate increase of the mobility with cooling (over a limited
$T$ range), similar to the case of "bulk" transport in the TOF
experiments. Does this mean that these devices disorder-free? Of
course, it does not. Since this characteristic time of the charge
release by shallow traps decreases exponentially with the
temperature, the shallow traps might be "invisible" at room
temperature. However, with cooling, they start dominating the
charge transport in single-crystal OFETs, as all the data obtained
so far suggest: the mobility drops exponentially, typically below
100 K. In this respect, the low-temperature $\mu(T)$ dependence
observed in the best single-crystal OFETs resemble the results of
TOF experiments with not-so-pure "bulk" crystals. The observed
exponential drop of $\mu(T)$ with cooling suggests that the
density of shallow traps near the surface of organic crystals in
single-crystal OFETs is still relatively large.

Though the low-temperature transport in single-crystal OFETs is,
most likely, disorder-dominated, it is worth considering
alternative explanations of a qualitative difference between the
results of OFET and TOF experiments. It is usually tacitly assumed
that the "surface" transport in OMCs is similar to the "bulk"
transport, which has been comprehensively studied in the 70s and
80s (see, e.g., \cite{Pope99}). However, the transport in organic
FETs differs from the "bulk" charge transport in the TOF
experiments at least in two important aspects. Firstly, one cannot
exclude \textit{a priori} the existence of some surface modes that
might affect the charge transport at the organic/insulator
interface. Secondly, even at a low gate voltage, the density of
charge carriers in a two-dimensional conducting channel of OFETs
exceeds by far the density of charges in the TOF experiments. The
interactions between polaronic excitations and their effect on the
charge transport in organic semiconductors is still an open issue
\cite{Capone03}. Thus, the experiments with single-crystal OFETs
pose new problems, which have not been addressed by the theory
yet.

Quantitative description of polaronic formation is a challenging
many-body problem. Band-type calculations for OMC provide guidance
in the search for new high-mobility materials, but do not have
predictive power since they ignore polaronic effects
\cite{Silinsh94}. For better characterization of the polaronic
states at the organic surface, it is crucial to go beyond the
\textit{dc} experiments and to measure directly the characteristic
energies involved in polaron formation. This information can be
provided by the infra-red spectroscopy of the conduction channel
in the single-crystal OFETs, a very promising direction of the
future research.

Though the current efforts are mostly focused on realization of
the intrinsic polaronic transport, equally important is to study
interaction of polarons with disorder. From the history of physics
of inorganic semiconductors, we know how important is this
research for better understanding of charge transport and device
characteristics. In organic conductors, small polarons interact
strongly with disorder, and this interaction is qualitatively
different from the interaction of electron-like excitations with
disorder in inorganic semiconductors. This research, which
encompasses such important issues as the spectroscopy of the
states in the band gap, mechanisms of polaron trapping by shallow
and deep traps, polaron scattering by defects in a
strongly-anisotropic medium will be critically important for both
basic science and applications. One of the aspects of this general
problem is the effect of stress on polaronic motion. Indeed, in
the field-effect experiments, it is difficult to avoid the
build-up of mechanical stress, caused by the difference in the
thermal expansion coefficients for the materials that form an
OFET. The stress in organic crystals might affect strongly the
overlap between electronic orbitals, the density of defects, and,
as the result, the temperature dependence of the mobility in the
FET experiments. More experiments with different FET structures
are needed to clarify this issue.

Realization of the intrinsic transport, especially at low
temperatures, requires better purification of starting material
and improvement of the crystal growth techniques. There is ample
room for improvement in this direction: the first experiments on
fabrication of single-crystal OFET did not use a powerful
zone-refinement technique, which has been proven to be very useful
in the TOF experiments \cite{Karl80,Karl85}. In fact, many of the
molecules used in FET experiments, including rubrene, can be zone
refined.

To understand better the intrinsic transport in organic
semiconductors and the relation between macroscopic electronic
properties and molecular packing, the experiments with OFETs based
on a broader class of organic crystals are required. Note that the
experiments with any type of organic crystals, regardless of the
magnitude of the \textit{intrinsic} mobility, are important - in
fact, only by exploring materials with different crystal
structures and, hence, a wide range of $\mu$, this problem can be
adequately addressed. However, realization of the coherent
band-like transport and improvement of the room-temperature
mobility in OFETs will be helped by the development of new organic
materials with a stronger overlap between electronic orbitals of
adjacent molecules. In this regard, the materials with a
non-planar molecular structure similar to that of rubrene
(conjugated chains with non-planar side groups) seem to be very
promising (see, e.g., \cite{Anthony02}). Recent experiments with
rubrene, which enabled a 10-fold increase of the OFET mobility,
illustrates how important is the effect of molecular packing on
the charge transport in organic crystals. To realize the enormous
potential of organic chemistry in "tailoring" new organic
compounds for organic electronics, combined efforts of chemists
and physicists are required.

The use of a broader variety of organic compounds for the OFET
fabrication may also offer the possibility to explore new
electronic phenomena. An interesting example is provided by the
metal-phthalocyanines (MPc's), a large class of materials with
isostructural molecules that differ by only one atom - the metal
element. This metal atom determines the electronic properties of
the molecules, including the spin S in the molecular ground state
\cite{Clack72,Liao01}. For instance, among the MPc's containing
$3d$ transition metal elements, MnPc has S$=3/2$ \cite{Awaga91},
FePc has S$=1$ \cite{Evangelisti02}, CoPc and CuPc have S$=1/2$
\cite{Liao01}. In FETs based on these materials, therefore, it may
be possible to induce and control electrostatically the magnetic
properties of organic materials, just by tuning the density of
mobile charge carrier that mediate interactions between the local
spins. Successful realization of 'magnetic' OFETs would open the
way for organic spintronics.

\section{Acknowledgments}

We are grateful to M. Jochemsen, N. Karl, V. Kiryukhin, T. M.
Klapwijk, Ch. Kloc, J. Niemax, J. Pflaum, V. Pudalov, J. Rogers,
and A. F. Stassen for useful discussions and help with
experiments. The work at Rutgers University was supported in part
by the NSF grant DMR-0077825 and the DOD MURI grant
DAAD19-99-1-0215. Work at Delft University was supported by the
"Stichting FOM" (Fundamenteel Onderzoek der Materie) and by NWO
via the Vernieuwingsimpuls 2000 program.


\begin{thebibliography}{99}
\bibitem{notetraps} The total number of the "bulk" charge traps per the unit
area of the conducting channel can be estimated as $N_t \times d$, where
$N_t \simeq 10^{15}$ cm$^{-3}$ is the bulk concentration of traps, and $d$
is the effective channel thickness ($\leq 10$ nm). The corresponding threshold
voltage, associated with the bulk traps in the studied devices, is therefore
very small ($< 0.1$ V).
\end{thebibliography}
\end{document}